\documentclass[11pt]{article}
\usepackage{epsf,amsfonts}
\usepackage{fontenc,indentfirst, delarray,amsfonts,amsmath,amssymb}
\usepackage{graphicx}

\DeclareGraphicsExtensions{ps,eps,ps.gz} \headheight 1.2cm
\makeatletter
\def\@citex[#1]#2{\if@filesw\immediate\write\@auxout
        {\string\citation{#2}}\fi
\def\@citea{}\@cite{\@for\@citeb:=#2\do
        {\@citea\def\@citea{,}\@ifundefined
        {b@\@citeb}{{\bf ?}\@warning
        {Citation `\@citeb' on page \thepage \space undefined}}
        {\csname b@\@citeb\endcsname}}}{#1}}
\newif\if@cghi
\def\cite{\@cghitrue\@ifnextchar [{\@tempswatrue
        \@citex}{\@tempswafalse\@citex[]}}
\def\citelow{\@cghifalse\@ifnextchar [{\@tempswatrue
        \@citex}{\@tempswafalse\@citex[]}}
\def\@cite#1#2{{\if@cghi\unskip$\null^{#1}$\else #1\fi\if@tempswa\typeout
        {warning: optional citation argument ignored: `#2'} \fi}}

\def\@biblabel#1{$\null^{#1}$}
\makeatother

\usepackage{epsf,amsfonts,amsmath,amssymb,fontenc,indentfirst,delarray,graphicx}
\newtheorem{lem}{Lemme}[section]

\newtheorem{prop}[lem]{Proposition}
\newtheorem{cor}[lem]{Corollary}
\newtheorem{defi}[lem]{Definition}

\newtheorem{che}{Check}

\newcommand{\norm}[1]{\left\lVert#1\right\rVert}
\newcommand{\abs}[1]{\lvert#1\rvert}

\newcommand{\scl}[2]{\langle#1,#2\rangle}
\newcommand{\suup}[1]{ \underset{#1}{\sup} }

\newcommand{\grad}[1]{\text{grad}\,#1}
\newcommand{\ket}[1]{\lvert#1\rangle}
\newcommand{\bra}[1]{\langle#1\lvert}
\newcommand{\tr}[1]{\text{Tr}(#1)}

\def\re{\text{Re}}
\def\im{\text{Im}}

\def\dm{\lp\begin{array}}
\def\fm{\end{array}\rp}
\def\dbb{\lb\begin{array}}
\def\fbb{\end{array}\rb}
\def\dbn{\left.\begin{array}}
\def\fbn{\end{array}\right.}

\def\lb{\left[}
\def\rb{\right]}
\def\lp{\left(}
\def\rp{\right)}

\def\m3{M_3 \lp \cc \rp}
\def\m2{M_2 \lp \cc \rp}

\def\mn{M_n \lp \cc \rp}

\def\cc{{\mathbb{C}}}
\def\rr{{\mathbb{R}}}
\def\nn{{\mathbb{N}}}
\def\zz{{\mathbb{Z}}}
\def\ii{{\mathbb{I}}}

\def\aa{{\cal A}}
\def\bb{{\cal B}}

\def\dd{{\cal D}}
\def\hh{{\cal H}}

\def\oo{{\cal O}}

\def\mm{{M}}
\def\pp{{\cal P}}

\def\cinf{C^{\infty}\lp\mm\rp}

\def\L2{L_2(\mm)}

\def\ot{\otimes}

\def\xox{{\xi}_x}

\def\xoz{{\zeta}_x}
\def\yoz{{\zeta}_y}

\def\ox{\omega_{\xi}}

\def\xo0{\omega^0_x}
\def\yo0{\omega^0_y}

\def\o0{\omega_0}

\def\xo0{x_\omega^0}
\def\yo0{y_\omega^0}

\def\nabc{\nabla_{\!\!\dot{c}_*}}
\def\nabm{\nabla_{\!\!\mu}}

\def\nabc{\nabla_{\!\!\dot{c}_*}}
\def\nabm{\nabla_{\!\!\mu}}
\begin{document}
\title{\bf {\Large Carnot-Carath\'eodory metric and gauge fluctuations in Noncommutative
Geometry}}
\author{Pierre Martinetti\\
Centre de Physique Th\'eorique,\\CNRS Luminy, case 907, F-13288
Marseille\\
{\it martinet@cpt.univ-mrs.fr}}
\date{\small\today} \maketitle

\begin{abstract}{\footnote{Paper published in Communications in Mathematical Physics.
The original publication is available at www.springerlink.com}}
Gauge fields have a natural metric interpretation in terms of
horizontal distance. This distance, also called
Carnot-Carath\'eodory or sub-Riemannian distance, is by definition
the length of the shortest horizontal path between points, that is
to say the shortest path whose tangent vector is everywhere
horizontal with respect to the gauge connection. In noncommutative
geometry all the metric information is encoded within the Dirac
operator $D$. In the classical case, i.e. commutative, Connes's
distance formula allows one to extract from $D$ the geodesic
distance on a Riemannian spin manifold. In the case of a gauge
theory with a gauge field A, the geometry of the associated
$U(n)$-vector bundle is described by the covariant Dirac operator
$D+A$. What is the distance encoded within this operator ? It was
expected that the noncommutative geometry distance $d$ defined by a
covariant Dirac operator was intimately linked to the
Carnot-Carath\'eodory distance $d_H$ defined by A. In this paper we
make this link precise, showing that the equality of $d$ and $d_H$
strongly depends on the holonomy of the connection. Quite
interestingly we exhibit an elementary example, based on a 2-torus,
in which the noncommutative distance has a very simple expression
and simultaneously avoids the main drawbacks of the Riemannian
metric (no discontinuity of the derivative of the distance function
at the cut-locus) and of the sub-Riemannian one (memory of the
structure of the fiber).
\end{abstract}

\section{Introduction}

Noncommutative geometry\cite{connes} enlarges differential geometry
beyond the scope of Riemannian spin mani\-folds and gives access, as
shown in various examples, to spaces obtained as the product of the
continuum by the discrete. It allows one to describe in a single and
coherent geometrical object the space-time of the Standard Model of
elementary particles \footnote{with massless neutrinos. Massive
Dirac neutrinos are easily incorporated in the model\cite{neutrinos}
as long as one of them remain massless. Otherwise more substantial
changes might be required.} coupled with Euclidean general
relativity \cite{spectral}. Specifically, the diffeomorphism group
of general relativity appears as the automorphism group of $\cinf$,
the algebra of smooth functions over a compact Riemannian spin
manifold $M$, while the gauge group of the strong and electroweak
interactions emerges as the group $U(\aa_I)$ of unitary elements of
a finite dimensional algebra $\aa_I$ (modulo a lift to the
spinors\cite{unim}). Remarkably, unitaries not only act as gauge
transformations but also acquire a metric significance via the
so-called {\it fluctuations of the metric}. This paper aims to study
in detail the analogy introduced in [\citelow{gravity}] between a
simple kind of fluctuations of the metric, those governed by a
connection $1$-form on a principal bundle, and the associated
Carnot-Carath\'eodory metric.

A noncommutative geometry consists in a {\it spectral triple}
$$\aa ,\; \hh,\; D$$
where $\aa$ is an involutive algebra, commutative or not, $\hh$ a
Hilbert space carrying a representation $\Pi$ of $\aa$ and $D$ a
selfadjoint operator on $\hh$. Together with a chirality operator
$\Gamma$ and a real structure $J$ both acting on $\hh$, they satisfy
a set of properties\cite{gravity} providing the necessary and
sufficient conditions for 1) an axiomatic definition of Riemannian
spin geometry in terms of commutative algebra  2) its natural
extension to the noncommutative framework. Points are recovered as
pure states $\pp(\aa)$ of $\aa$, in analogy with the commutative
case where
\begin{eqnarray}
&\pp(\cinf) \simeq \mm& \\
&\label{etatcomm} \omega_x(f)=f(x)&
\end{eqnarray}
for any pure state $\omega_x$ and smooth function $f$. A distance
$d$ between states $\omega$, $\omega'$ of $\aa$ is defined by
\begin{equation}
\label{distance} d(\omega, \omega') \doteq \suup{a\in\aa}\left\{
\abs{\omega(a) - \omega'(a)}\,;\; \norm{[D,\Pi(a)]}\leq 1\right\}
\end{equation}
where the norm is the operator norm on $\hh$. In the
commutative case,
\begin{equation}
\label{continu} \aa_E = \cinf,\; \hh_E=L_2(M,S),\; D_E=
-i\gamma^\mu\partial_\mu
\end{equation}
with $\hh_E$ the space of square integrable spinors and $D_E$ the
ordinary Dirac operator of quantum field theory, $d$ coincides with
the geodesic distance defined by the Riemannian structure of $\mm$.
Thus (\ref{distance}) is a natural extension of the classical
distance formula, all the more as it does not involve any notion
ill-defined in a quantum framework such as the trajectory between
points.

Carnot-Carath\'eodory metrics (or sub-Riemannian
metrics)\cite{montgomery} are defined on manifolds $P$ equipped with
a {\it horizontal distribution}, that is to say a (smooth)
specification at any point $p\in P$ of a subspace $H_p P$ of the
tangent space $T_pP$. The Carnot-Carath\'eodory distance $d_H$
between $p$ and $q$ is the length of the shortest path $c$ joining
$p$ and $q$ whose tangent vector is everywhere horizontal,
\begin{equation}
\label{dcc}
d_H(p,q) = \underset{\dot{c}(t)\in H_{\!c(t)}\! P}{\text{Inf}}\; \int_0^1 \norm{\dot{c}(t)} dt.
\end{equation}
 If there is no horizontal path from $p$ to $q$ then $d_H(p,q)$ is infinite. Any point at finite distance from $p$ is said {\it accessible}
\begin{equation}
\label{acc}
\text{Acc}(p)\doteq \{q\in P;\; d_H(p,q) < +\infty\}.
\end{equation} Most often the norm in the
 integrand of (\ref{dcc}) comes
from an inner product in the horizontal subspace. The latter can be
obtained in (at least) two ways: either by restricting to $HP$ a
Riemannian structure of $TP$ or, when $P\overset{\pi}{\rightarrow}
M$ is a fiber bundle
 with a connection, by pulling back the Riemannian structure $g$ of $M$. In the
 latter case the horizontal distribution is the kernel of the connection
$1$-form and any horizontal vector has norm
\begin{equation}
\label{normeh}
\norm{u} \doteq \norm{\pi_*(u)} = \sqrt{g(\pi_*(u), \pi_*(u))}.
\end{equation}
Note that (\ref{dcc}) provides $P$ with a distance although $P$ may
not be a metric manifold, only $M$ is asked to be Riemannian.

 By taking the product of a Riemannian geometry (\ref{continu}) by
 a spectral triple with finite dimensional $\aa_I$, one obtains as
 pure state
 space a $U(\aa_I)$-bundle $P$ over $M$. A
 connection on $P$ then not only defines
 a Carnot-Carath\'eodory distance $d_H$ but also, via the process of fluctuation of the
 metric recalled in section \ref{fluctuation}, a distance $d$ similar to
 (\ref{distance}) except that the ordinary Dirac operator $D$ is replaced by
 the covariant differentiation operator associated to
 the connection-1 form. In
section \ref{composanteconnes} we compare the connected components
for these two distances: while a connected component for $d_H$ is
also connected for $d$, a connected component of $d$ is not
necessarily connected for $d_H$. We investigate the importance of
the holonomy group on this matter. In section \ref{flatcase} we show
that the two distances coincide when the holonomy is trivial. In the
non-trivial case we work out some necessary conditions on the
holonomy group that may allow $d$ to equal $d_H$. In section
\ref{contrex} we treat in detail a simple low-dimensional example in
which each of the connected components of $d_H$ is a dense subset of
a two dimensional torus $\mathbb{T}$. As a main result of this paper
we show in section \ref{interpretation} that while the
Carnot-Carath\'eodory metric forgets about the fiber bundle
structure of $\mathbb{T}$, the noncommutative metric deforms it in a
quite intriguing way: from a specific intrinsic point of view, the
fiber acquires the shape of a cardioid. Hence the classical
$2$-torus inherits a metric which is "truly" noncommutative in the
sense that it cannot be described in (sub)Riemannian nor discrete
terms. This is, to our knowledge, the novelty of the present work.
\\

\noindent{\bf Notations and conventions:}
\begin{itemize}
\item $\mm$ is a Riemannian compact spin manifold of dimension $m$
without boun\-dary. Cartesian coordinates are labeled by Greek
indices $\mu, \nu$ and we use Einstein summation over repeated
indices in alternate positions (up/down).

\item  $\pp(\aa)$ denotes the set of pure states of $\aa$
(positive, linear applications from $\aa$ to $\cc$, with norm $1$
and that do not decompose as a convex combination of other states).
Throughout this paper we deal with a pure state space which is  a
trivial bundle $P$ over $M$, with fiber $\cc P^{n-1}$. An element of
$P$ is written $\xi_x$ where $x$ is a point of $M$ and $\xi\in\cc P
^{n-1}$.

\item Most of the time we omit the symbol $\Pi$ and it should
be clear from the context whether $a$ means an element of $\aa$ or
its representation on $\hh$. Unless otherwise specified a bracket
denotes the scalar product on $\cc^n$.

\item We use the result of [\citelow{finite}] according to which the
supremum in (\ref{distance}) can be sought on positive elements of
$\aa$.
\end{itemize}
\newpage
\section{\label{fluctuation} Fluctuations of the metric}

In noncommutative geometry a connection on a geometry $(\aa, \hh,
D)$ is defined via the identification of $\aa$ as a finite
projective module over itself (i.e. as the noncommutative equivalent
of the sections of a vector bundle via the Serre-Swan
theorem)\cite{gravity}. It is implemented by replacing $D$ with a
{\it covariant operator}
\begin{equation}
\label{diraccov} D_A \doteq D + A + JAJ^{-1}
\end{equation} where $J$ is the real structure and $A$ is a selfadjoint element of the set
$\Omega^1$ of 1-forms
\begin{equation}
\Omega^1 \doteq  \left\{a^i[D,b_i]\;;\; a^i,b_i\in\aa\right\}.
\end{equation}
Only the part of $D_A$ that does not obviously commute with the
representation, namely
\begin{equation}
\label{tildeD} \mathcal{D} \doteq D + A,
\end{equation}
enters in the distance formula (\ref{distance}) and induces a
so-called {\it fluctuation of the metric}. In the following we
consider almost commutative geometries obtained as the product of
the continuous - external - geometry (\ref{continu}) by an internal
geometry $\aa_I,\; \hh_I,\; D_I$. The product of two spectral
triples, defined as
\begin{equation}
\label{tripletprod}
 \aa = \aa_E\ot\aa_I,\; \hh= \hh_E\ot
\hh_I,\; D= D_E \ot \ii_I + \gamma^5\ot D_I
\end{equation}
where $\ii_I$ is the identity operator of $\hh_I$ and $\gamma^5$
the chirality of the external geometry, is again a spectral
triple. The corresponding 1-forms are\cite{kt,vh}
$$
-i\gamma^\mu f_\mu^i \ot m_i + \gamma^5 h^j\ot n_j$$ where
$m_i\in\aa_I$, $h^j, f_\mu^i\in\cinf$, $n_j\in\Omega_I^1$.
Selfadjoint $1$-forms decompose into an  $\aa_I$-valued skew-adjoint
1-form field over $\mm$, $A_\mu\doteq f^i_\mu m_i$, and an
$\Omega^1_I$-valued selfadjoint scalar field $H\doteq h^j n_j.$

When the internal algebra $\aa_I$ has finite dimension, $A_\mu$
takes values in the Lie algebra of unitaries of $\aa$ and is called
the {\it gauge part} of the fluctuation. In [\citelow{kk}] we have
computed the noncommutative distance (\ref{distance}) for a scalar
fluctuation only ($A_\mu = 0$). In [\citelow{gravity}] the distance
is considered for a pure gauge fluctuation ($H = 0$) obtained from
the internal geometry
$$\aa_I =
M_n(\cc),\quad \hh_I = M_n(\cc),\quad D_I = 0,$$
that is to say
\begin{equation}
\label{tildedeux}
 {\cal{D}} = -i\gamma^\mu (\partial_\mu \ot \ii_I  + \ii_E\ot A_\mu).
\end{equation}
$\aa_E$ being nuclear, the set of pure states of
\begin{equation}
\label{atot}
\aa =\cinf\ot \mn
= C^{\infty}(M,\mn)
\end{equation}
 is \cite{kadison}
 $\pp(\aa) \simeq \pp(\aa_E) \times \pp(\aa_I)$, where
 $\pp(\aa_I)$ is the projective
 space $\cc P^{n-1}$,
 \begin{equation}
 \label{evaluationun}
 \omega_\xi(m) = \scl{\xi}{m\xi} = \tr{s_\xi\, m}
 \end{equation}
for $m\in\aa_I, \xi\in\cc P^{n-1}$ and $s_\xi$ the support of
$\ox$. The evaluation of $\xox \doteq
 (\omega_x, \omega_\xi)$ on $a=f^i\ot m_i\in\aa$ reads
\begin{equation}
\label{evaluationdeux}
 \xi_x(a)= \tr{s_\xi\, a(x)}
 \end{equation}
where
\begin{equation}
\label{adex}
 a(x) \doteq f^i(x)\ot m_i.
\end{equation}
Hence $\pp(\aa)$ is a trivial bundle
$$P\overset{\pi}
{\rightarrow}\nolinebreak M$$ with fibre $\cc P^{n-1}$.

The gauge potential $A_\mu$ defines both a horizontal distribution
$H$ on $P$, with associated Carnot-Carath\'eodory metric $d_H$, and
a noncommutative metric $d$ given by formula (\ref{distance}) with
$\dd$ substituted for $D$. In the case of a zero connection, $\dd =
D_E$ and $d$ is the geodesic distance on $M$. Indeed the commutator
norm condition $\norm{[D_E,f]}\leq 1$ forces the gradient of $f$ to
be smaller than $1$, so that
\begin{equation}
\label{cascomm}
d(\omega_x,\omega_y) = \underset{\norm{\grad{f}}\leq 1}{\text{sup}}\abs{f(x) - f(y)}
\leq \int_0^1 \norm{\dot{c}(t)}dt = d_\text{geo}(x,y)
\end{equation}
where $c$, $c(0)=x$, $c(1)=y$ is a minimal geodesic from $x$ to $y$.
One then easily checks that this upper bound is attained by the
function
\begin{equation}
\label{naifl}
L(z)\doteq d_\text{geo}(x,z) \quad \forall z\in M
\end{equation}
(or more precisely by a sequence of smooth functions converging to
the continuous function $L$). As we shall see in the following
section, in the case of a non-zero connection,
 one obtains without difficulty a result similar to (\ref{cascomm}) with $d_H$
playing the role of $d_\text{geo}$ (cf eq. (\ref{ineq1}) below).
However, except in some simple cases studied in section
\ref{flatcase}, $d_H$ is not the least upper bound and there is no
simple equivalent to the function $L$. In fact the main part of this
paper, especially section \ref{contrex}, is devoted to building the
element $a\in\aa$ that realizes the supremum in the distance
formula.

\section{Connected components \label{composanteconnes}}

We say that two pure states $\xox$, $\yoz$ are connected for $d$
if and only if $d(\xox, \yoz)$ is finite.
{\prop  \label{lemme1} For any $\xox$ in $P$, $\text{Acc}(\xox)$
is connected for $d$.
}
\\

\noindent {\it Proof.} The result is obtained by showing that for any
$\yoz\in\text{Acc}(\xox)$,
\begin{equation}
\label{ineq1}
d(\xox,\yoz)\leq d_H(\xox,\yoz).
\end{equation} Let
us start by recalling how to transfer the covariant
derivative\cite{kob} of a section $V$ of $P$,
$$\nabla_\mu V = \partial_\mu V + A_\mu V,
$$
to the algebra $\aa$. Given $a\in\aa$, the evaluation
(\ref{evaluationdeux}) is the diagonal of the sesquilinear form
defined fiberwise on the vector bundle
$P'\overset{\pi'}\rightarrow M$ with fiber $\cc^n$,
\begin{equation}
\label{sesq}
(W_x',V_x') \mapsto \scl{W'_x}{a(x)V_x'}
\end{equation}
for $W_x',V_x'\in \pi'^{-1}(x)$. Accordingly, as a $\cinf$-module,
we view $\aa$ as the sections of the bundle $P''$ of rank-two
tensors on $M$
$$a = a_{ij}\; \overline{e^i} \ot e^j
$$
with values in ${\overline{T^*\cc^n}}\ot T^*{\cc^n}$. Here $\{e^i\}$
is the dual of the canonical basis $\{e_i\}$ of $T\cc^n\simeq\cc^n$
and $\{\overline{e^i}\}$ its complex conjugate
$$\overline{e^i}(V) = \overline{V^i}\;\;\text{for}\;\; V = V^i e_i\in\cc^n.$$
The covariant derivative of $P$ then naturally extends to $P''$
\footnote{
$\left\{\begin{array}{c} \nabm e^i = -A_{\mu k}^i e^k\\
\nabm \overline{e^i} = -\overline{A_{\mu k}^i}\, \overline{e^k}
\end{array}\right.$
hence
$\nabm a \doteq \nabm (a_{ij}\, \overline{e^i})\ot e^j + a_{ij}\, \overline{e^i}\ot
\nabm e^j = (\partial_\mu a_{ij} + [A,a]_{ij}) \overline{e^i}\ot e^j.$}
\begin{equation}
\label{covsec}
\nabm a = \partial_\mu a + [A_\mu,a].
\end{equation}

Let us fix a horizontal curve of pure states $c(t)$, $t\in[0,1]$,
between $\xox$ and $\yoz$ as defined in (\ref{evaluationdeux}). Let
$\lp \pi, V\rp$ be a trivialization in $P$ such that
\begin{equation}
\label{loctriv} \pi(\xi_x) = x,\;\, V(\xi_x) =  \xi \quad
\quad \pi(\zeta_y) = y,\;\, V(\zeta_y) = \zeta
\end{equation}
and define
$$V(t) \doteq V(c(t)).$$
$c$ is the horizontal lift starting at
$\xox$ of the curve
$$c_*(t)\doteq \pi(c(t))
$$
lying in $M$ and tangent to
\begin{equation}
\pi_*(\dot{c}) = \dot{c}_* = \dot{c}_*^\mu\partial_\mu.
\end{equation}
 Writing $s(t)$ for the support of the pure state $\omega_{V(t)}$, the curve $t\mapsto s(t)$ is
 horizontal in $P''$ in the sense of the covariant derivative (\ref{covsec})
{\footnote{in Dirac notation $c$ horizontal in $P$ is written
$\dot{\ket{V}} + \dot{c}^\mu A_\mu \ket{V} = 0$. By simple
manipulations $\left\{
\begin{array}{c} \dot{\ket{V}}\bra{V}\, +\, \dot{c}^\mu A_\mu
\ket{V}\bra{V}\, = 0 \\\ket{V}\dot{\bra{V}} - \ket{V}\bra{V}
\dot{c}^\mu A_\mu = 0\end{array}\right. ,$ hence $\dot{s} =
\ket{V}\dot{\bra{V}} + \dot{\ket{V}}\bra{V} =
\dot{c}^\mu[\ket{V}\bra{V}, A_\mu] = \dot{c}^\mu[s,A_\mu]$.}}
\begin{equation}
\label{transport}
\nabc s \doteq \dot{c}_*^\mu \, \nabm s = 0.
\end{equation}
Let us associates to any $a\in\aa$ its evaluation $f$ along $c$,
\begin{equation}
\label{fonction}
 f(t)\doteq \tr{s(t)a(c_*(t))},
 \end{equation}
 whose derivative with respect to $t$ is easily computed using (\ref{transport})
\begin{equation}
\label{derivf}
\dot{f} = \tr{s \, \nabla_{\dot{c}_*}a}.
\end{equation}
At a given $t$ the Cauchy-Schwarz inequality yields the bound
 \begin{equation}
 \label{cs}
\abs{\dot{f}(t)}\leq \norm{df_{\lvert t}}\norm{\dot{c}_*(t)}
\end{equation}
where $df$ is the $1$-form on $c_*$ with components
\begin{equation}
\label{deltaf}
\partial_\mu f = \tr{s \nabm a}.
\end{equation}
$s[{\cal{D}},a]s$ evaluated at some $c_*(t)$ is an $n'\times n$
square matrix ($n' = \text{dim}\, \hh_E$ is the dimension of the
spin representation),
\begin{equation} \label{evalun}
 s[{\cal{D}},a]s = -i\gamma^\mu \ot s (\nabla_\mu a)s = -i\gamma^\mu \partial_\mu f \ot s,
\end{equation}
with norm $\norm{df_{\lvert t}}$. Therefore
\begin{equation}
\label{normedirac}
 \displaystyle
 \norm{df_{\lvert t}}\leq
 \suup{x\in
M}\norm{[{\cal{D}},a]_{| x}} = \norm{[{\cal{D}},a]}
\end{equation}
so, as soon as $\norm{[\dd,a]} \leq 1$,
\begin{equation}
\displaystyle \abs{\xox(a) - \yoz(a)} = \abs{{\int_0}^1
\dot{f}(t)\, dt\,} \leq {\int_0}^1 \norm{\dot{c}_*(t)}{dt},
\label{integun}
\end{equation}
which precisely means $ d(\xox,\yoz) \leq
d_H(\xox,\yoz)$.\hfill$\blacksquare$
\\

It would be tempting to postulate that $d$ and $d_H$ have the same
connected components. Half of this way is done in the proposition
above. The other half would consist in checking that $d$ is infinite
as soon as $d_H$ is infinite. However this is, in general, not the
case. It seems that there is no simple conclusion on that matter
since we shall exhibit in section \ref{contrex} an example in which
some states that are not in $\text{Acc}(\xox)$ are at finite
noncommutative distance from $\xox$ whereas others are at infinite
distance. The best we can do for the moment is to work out
(Proposition \ref{toplem} below) a sufficient condition on the
holonomy group associated to the connection $A_\mu$ that guarantees
the non-finiteness of $d(\xox,\yoz)$ for $\yoz\notin
\text{Acc}(\xox)$. We begin with the following elementary lemma.

{\lem \label{infini} $d(\xox, \yoz)$ is infinite if and only if
there is a sequence $a_n\in\aa$ such that
\begin{equation}
\label{condinf} \underset{n\rightarrow +\infty}{\text{lim}}
\norm{[D,a_n]} \rightarrow 0,\quad \underset{n\rightarrow
+\infty}{\text{lim}} \abs{\xox(a_n)-\yoz(a_n)} = +\infty.
\end{equation}}

\noindent{\it Proof.} The point is to show that from a sequence
$a_n$ satisfying
$$\norm{[D,a_n]} \leq 1\; \forall n\in\nn,\quad \underset{n\rightarrow
+\infty}{\text{lim}} \abs{\xox(a_n)-\yoz(a_n)} = +\infty
$$ one can extract a sequence $\tilde{a}_n$ satisfying (\ref{condinf}).
This is done by considering $$\tilde{a}_n \doteq
\frac{a_n}{\sqrt{\abs{\xox(a_n)-\yoz(a_n)}}}.$$
\hfill$\blacksquare$

{\prop \label{toplem} Let $\xi,\zeta\in\cc P^{n-1}$. If there
exists a matrix $M\in\mn$ that commutes with the holonomy group at
$x$, $\text{Hol}(x)$, and such that
\begin{equation}
\label{trace} \tr{s_\xi M}\neq\tr{s_\zeta M}, \end{equation}
 then
$d(\omega,\omega')=+\infty$ for any $\omega\in\text{Acc}(\xox)$,
$\omega'\in \text{Acc}(\xoz)$.}
\\

\noindent{\it Proof.} The proof is a restatement of a classical
result (cf  [\citelow{lichne}] p.113) according to which an
element of $\aa$ invariant under the adjoint action of the
holonomy group is a parallel tensor, that is to say $\nabm a = 0$
in all directions $\mu$. We detail this point in the following for
the sake of completeness.

From now on we fix a trivialization $(\pi, V)$ on $P=\pp(\aa)$. Recall that
given a curve from $c_*(0)=x$ to $c_*(1)=y\in\mm$, the end point
of the horizontal lift $c$ of $c_*$ with initial condition
$c(0)=(x,\xi)$ is $c(1) = (y, U_{c_*}(1)\xi)$ where
$$U_{c_*}(t) = P \displaystyle \exp (-{ \int_{c_*(t)} {A_\mu dx^\mu}})$$
($P$ is the time-ordered product) is the solution of
\begin{equation}
\label{parral1} \dot{U} = -\dot{c}^\mu A_\mu U.
\end{equation}
In the following we write $U_{c_*}$ for $U_{c_*}(1)$.
Let $M\in\mn$ commute with $\text{Hol}(x)$. Define $a_M\in\aa$ by
$$
a_M(x)\doteq M $$ and for any $y\in\mm$,
\begin{equation}
\label{am} a_M(y) \doteq U_{c_*} a_M(x)U_{c_*}^*
\end{equation}
where $c_*$ is a curve joining $x$ to $y$. One checks that
$a_M(y)$ commutes with any $V_l\in\text{Hol}(y)$ since
$$
V_l \, a_M(y) V_l^* = U_{c_*}U_{c_*}^* V_l U_{c_*} a_M(x)U_{c_*}^*
V_l^*U_{c_*}U_{c_*}^* = a_M(y)
$$
where we use that $U_{c_*}^* V_l U_{c_*}$ belongs to
$\text{Hol}(x)$. Hence (\ref{am}) uniquely defines $a_M(y)$ since
parallel transporting $a_M(x)$ along another curve $c'_*$ yields
$${a'}_M(y) = U_{c'_*} U_{c_*}^* a_M(y) U_{c_*} U_{c'_*}^* = a_M(y)$$
where we used that $U_{c_*} U_{c'_*}^*\in\text{Hol}(y)$. Using
(\ref{parral1}) one explicitly checks that
$$\nabc \!a_M = 0.$$
Since this is true for any curve $c_*$, $a_M$ is parallel so
$$[\dd,a_M]= 0.$$
Now (\ref{trace}) means that $\xox(a_M) - \xoz(a_M) \neq 0$, hence
$d(\xox,\xoz) = +\infty$ by lemma \ref{infini}, and the result
follows by the triangle inequality. \hfill $\blacksquare$
\\

\noindent Proposition above only provides sufficient conditions.
Whether they are necessary, i.e. whether from
$d(\xox,\yoz)=+\infty$ on can build a matrix $M$ that commutes
with the holonomy group and do not cancel the difference of the
states is an open question. Lemma \ref{infini} suggests that to
any infinite distance is associated a tensor that commutes with
the Dirac operator. Moreover it is not difficult to show that any
parallel tensor commutes with the holonomy group. Therefore the
question is: are the parallel tensors the only ones that commute
with $D$ ? For the time being the answer is not clear to the author.

To close this section, let us mention a situation in which the two
metrics have the same connected components.

 {\cor
 \label{span}
 If for a given
$\xox\in P$ the vector space
$$\hh_{hol} \doteq \text{Span}\{U\xi\, ;\, U\in \text{Hol}(x)\}$$
has dimension $h < n$, then $\text{Acc}(\xox)$ is the connected component of $\xox$ for
 $d$.}
\\

\noindent{\it Proof.} In an orthonormal basis $\{\bb_{hol}, \bb\}$
of $\cc^n$ with $\bb_{hol}$ a basis of $\hh_{hol}$, $\text{Hol}(x)$
is block represented, so
$$M = \dm{cc}  0& 0 \\ 0 &\ii_{n-h}\fm$$
commutes with $\text{Hol}(x)$. Moreover $\tr{s_\xi M} = 0$. On the
contrary for any $\xoz\notin \text{Acc}(\xox)$, the rank one
projector $s_\zeta$ does not project on $\hh_{hol}$ so $\tr{s_\zeta
M} \neq 0$. Therefore, by Proposition \ref{toplem}, $d(\xox,\yoz)$
is infinite for any $\yoz\notin\text{Acc}(\xox)$, hence the result
by Proposition \ref{lemme1}.\hfill $\blacksquare$

\section{Flat case versus holonomy constraints\label{flatcase}}

The preceding section suggests that the two metrics defined by a
connection on the pure state space $P$ of the algebra (\ref{atot}),
the Carnot-Carath\'eodory distance $d_H$ and the noncommutative
distance $d$, do not coincide. It is likely that the two metrics do
not have the same connected components as soon as the conditions of
Proposition \ref{toplem} are not fulfilled. However nothing forbids
$d$ from equalling $d_H$ on each connected component of $d$. We
already know that $d\leq d_H$ so to obtain the equality it would be
enough to exhibit one positive $a\in\aa$ (or a sequence of elements
$a_n$) satisfying the commutator norm condition as well as
\begin{equation}
\label{maxi} \xox(a) - \yoz(a) = d_H(\xox,\yoz).
\end{equation}
The existence of such an $a$ strongly depends on the holonomy of the
connection: when the latter is trivial, e.g. by the Ambrose-Singer
theorem when the connection is flat and $M$ simply connected, then
the two metrics are equal, as shown below in Proposition
\ref{flatrat}. When the holonomy is non-trivial, we work out in
Proposition \ref{ososip} some necessary conditions on the shortest
path that may forbid $d$ from equalling $d_H$.

{\prop \label{flatrat} When the holonomy group
reduces to the identity, $d=d_H$ on all $P$.}
\\

\noindent {\it Proof.} For $\yoz\notin\text{Acc} (\xox)$, Corollary
\ref{span} yields $$d(\xox,\yoz) = +\infty = d_H (\xi_x, \zeta_y).$$
Thus we focus on the case $\yoz\in\text{Acc} (\xox)$. By Cartan's
structure equation the horizontal distribution defined by a
connection with trivial holonomy is involutive, which means that the
set of horizontal vector fields is a Lie algebra for the Lie bracket
inherited from $TP$. Equi\-valently (Frobenius theorem) the bundle
of horizontal vector fields is integrable. Hence $\text{Acc}(\xi_x)$
is a subma\-nifold of $P$, call it $\Xi$, such that $Tp\Xi = H_pP$
for any $p\in\Xi$. For any $z\in \mm$ there is at least $1$ point in
the intersection
$$\pi^{-1}(z)\cap\Xi$$
(e.g. the end point of the horizontal lift, starting at $\xox$, of
any curve from $x$ to $z$) and only one (otherwise there would be a
horizontal curve joining two distinct points in the fiber,
contradicting the triviality of the holonomy). In other words all
the horizontal lifts starting at $\xox$ of curves joining $x$ to $z$
have the same end point, call it $\sigma(z)$, and the application
$$\sigma: z \mapsto \pi^{-1}(z)\cap\Xi$$
defines a smooth section of $P$. Hence
$$\Xi = \sigma(\mm).
 $$
Note that $\zeta_y = \sigma(y)$ is the only point in the fiber over
$y$ which is at finite distance from $\xox=\sigma(x)$. Considering
the horizontal lift of the Riemannian geodesic from $x$ to $y$, it
turns out that $d_H$ on $\Xi$ coincides with the geodesic distance
$d_{\text{geo}}$ on $\mm$. The sequence of elements $a_n$ we are
looking for in (\ref{maxi}) is a sequence approximating the
continuous $\m2$-valued function
\begin{equation}
\label{solplate} L\ot \ii
\end{equation}
where $L$ is the geodesic distance function (\ref{naifl}).\hfill$\blacksquare$

The difficulty arises when the shortest horizontal curve $c$ does
not lie in a horizontal section. This certainly happens when the
connection is not flat and/or $M$ not simply connected. As soon as
the holonomy is non-trivial, different points $\xox$, $\xoz$ on the
same fiber can be at finite non-zero Carnot-Carath\'eodory distance
from one another although the Riemannian distance of their
projections vanishes. The question reduces to finding the equivalent
of the element (\ref{solplate}) in the closure of $\aa$ that attains
the supremum in (\ref{maxi}). A natu\-ral candidate to play the role
of the function $L$ in the case of a non-trivial holonomy is the
fiber-distance function which associates to any $z\in M$ the length
of the shortest horizontal path joining $\xox$ to some point in
$\pi^{-1}(z)$. When the holonomy is trivial this function precisely
coincides with $L$. However there is no natural candidate to play
the role of the identity matrix in (\ref{solplate}). Possibly one
might determine by purely algebraic techniques which element $a$ of
$\aa$ realizes the supremum in the distance formula. The best
approach we found for the moment is to work out, Proposition
\ref{ososip}, some conditions between the matrix part of $a$ and the
self-intersecting points of $c_*$ that are necessary for $d$ to
equal $d_H$.

\begin{figure}[hh]
\begin{center}
\mbox{\rotatebox{0}{\scalebox{.7}{\includegraphics{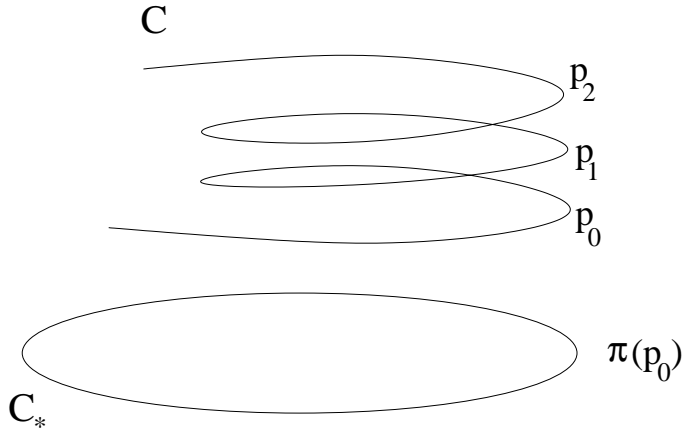}}}}
\end{center}
\caption{ \label{ellipse} An ordered sequence of self-intersecting
points, with $p_i=c(t_i)$.}
\end{figure}

{\defi
\label{osip}
 Given a curve $c$ in a fiber bundle with horizontal
distribution $H$, we call a c-ordered sequence of
$K$ self-intersecting points at $p_0$ a set of at least two elements
$\{c(t_0), c(t_1),...,c(t_K)\}$  such that
$$\pi(c(t_i)) = \pi(c(t_0)),\quad d_H(c(t_0), c(t_i)) >  d_H(c(t_0),
c(t_i))$$ for any $i=1,... ,K$ (Figure \ref{ellipse}).
}
{\lem \label{ossip} Let $\xox$, $\yoz$ be two points in $P$ such
that $d(\xox,\yoz) = d_H(\xox,\yoz)$. Then for any $c(t)$ belonging to a
minimal horizontal curve $c$ between $c(0)=\xox$ and $c(1)=\yoz$,
\begin{equation}
\label{unter} d(\xox,c(t))= d_H(\xox,c(t)).
\end{equation}
Moreover, for any such curve there exists an element $a\in\aa$ (or a
sequence $a_n$) such that
\begin{equation}
\label{deuxter}
 \xi_t(a) = d_H(\xox,c(t))
\end{equation} for any $t\in[0,1]$, where $\xi_t$ denotes $c(t)$ viewed
as a pure state of $\aa$.
}
\\

\noindent{\it Proof.} We write the proof assuming that the supremum
in the distance formula is attained by some $a\in\aa$. In case the
supremum is not reached, the proof is identical using a sequence
$\{a_n\}$. Assume $a$ does satisfies the commutator norm condition
as well as (\ref{maxi}). Let us parameterize $c$ by its length
element $\tau$ and use "dot" for the derivative $\frac{d}{d\tau}$.
The function $f(t)=\xi_t(a)$ defined by (\ref{fonction}) has
constant derivative along $c_*$. Indeed (\ref{maxi}) reads
\begin{equation}
\label{lambda}
 \int_0^\Lambda \dot{f}(\tau)d\tau = \Lambda
\end{equation}
where $\Lambda = d_H(\xox,\yoz)$. Since $\norm{\dot{c}_*(\tau)}=1$
for any $\tau\in[0,\Lambda]$, (\ref{cs}) and (\ref{normedirac})
forbid $\abs{\dot{f}(\tau)} $ from being greater than $1$. Hence
\begin{equation}
\label{fpoint} \dot{f}(\tau) = 1
\end{equation}
for almost every $\tau$. Thus for any $\lambda\leq\Lambda$,
\begin{equation}
\label{lambdabis}
 \int_0^{\lambda} \dot{f}(\tau)d\tau = \lambda
\end{equation}
which reads
\begin{equation}
\xi_\lambda(a) - \xox(a) = \lambda = d_H(\xox,\xi_\lambda).
\end{equation}
Hence (\ref{unter}) by Proposition \ref{lemme1}, and (\ref{deuxter})
by considering $\tilde{a}\doteq a-\xox(a)$. \hfill $\blacksquare$
%
\\

\noindent Applying lemma \ref{ossip} to the self-intersecting
points defined in \ref{osip} one obtains the announced necessary
conditions for $d$ to equal $d_H$.

{\prop \label{ososip} The noncommutative distance between two points
$\xox$, $\yoz$ in $P$ can equal the Carnot-Carath\'eodory one only
if there exists a minimal horizontal curve $c$ between $\xox$ and
$\yoz$ such that
 there exists  an element $a\in\aa$, or a sequence of
elements $a_n$, satisfying the commutator norm condition as well
as
\begin{equation}
\xi_{t_i}(a) = d_H(\xox, c(t_i))\;\text{ or }\; \underset{n\rightarrow
\infty}{\text{lim}} \xi_{t_i}(a_n) = d_H(\xox, c(t_i))
\end{equation}
for any $\xi_{t_i} = c(t_i)$ in any c-ordered sequence of self-intersecting
points.}
\newline

Given a sequence of $K$ self-intersecting points at $p$, Proposition
\ref{ososip} puts $K+1$ condition on the $n^2$ real components of
the selfadjoint matrix $a(\pi(p))$. So it is most likely that a
necessary condition for $d(\xox,\yoz)$ to equal $d_H(\xox,\yoz)$ is
the existence of a minimal horizontal curve between $\xox$ and
$\yoz$ such that its projection does not self-intersect more than
$n^2-1$ times. We will refine this interpretation in the example of
the next section. From a more general point of view it is not clear
how to deal with such a condition in the framework of sub-Riemannian
geometry{\footnote{Thanks to R. Montgomery\cite{montgomery} for
illuminating discussions on this matter.}}. It might be possible
indeed that in a manifold of dimension greater than $3$ one may, by
smooth deformation, reduce the number of self-intersecting points of
a minimal horizontal curve. But this is certainly not possible in
dimension $2$ or $1$. In particular, when the basis is a circle
there is only one horizontal curve $c$ between two given points, and
it is not difficult to find a connection such that $c_*$
self-intersects infinitely many times. This is what motivates the
following example.

\section{The example $C^\infty(S^1)\ot \m2$\label{contrex}}

 Let us summarize our comparative analysis of $d$ and $d_H$.
 When the holonomy is trivial
 the two distances are equal by proposition \ref{flatrat}. When the holonomy is non-trivial we have both:

- a sufficient, but maybe not necessary, condition (Corollary
\ref{span}) that guarantees the two distances have the
 same connected components,

 -a necessary condition
 (Proposition \ref{ososip}) for the two distances to
 coincide on a given connected component.

 \noindent These two
 conditions do not seem to be related:
 writing $Q^i$ and $Q_H^i$ the connected components of
 $d$ and $d_H$ respectively, it is likely that in some situations
 $Q^i=Q_H^i$ for some $i$ although $d$ differs from $d_H$ on $Q^i$, or on the contrary
 $Q_H^i \varsubsetneq Q^i$
 but $d = d_H$ on $Q_h^i$. In the present section
 we exhibit a simple low-dimensional example in which the $Q^i$'s are two dimensional
 tori (Proposition \ref{tori}) and the $Q_H^i$'s are dense subsets. $d$ coincides with
$d_H$ only on some part of
 $Q_H^i$ (Corollary \ref{erreur}).
 The present section is technical
 and deals with the exact
computation of the noncommutative distance (Proposition \ref{s1}).
Interpretation and discussion are postponed to the following
section.

\begin{figure}[hh]
\begin{center}
\mbox{\rotatebox{0}{\scalebox{.6}{\includegraphics{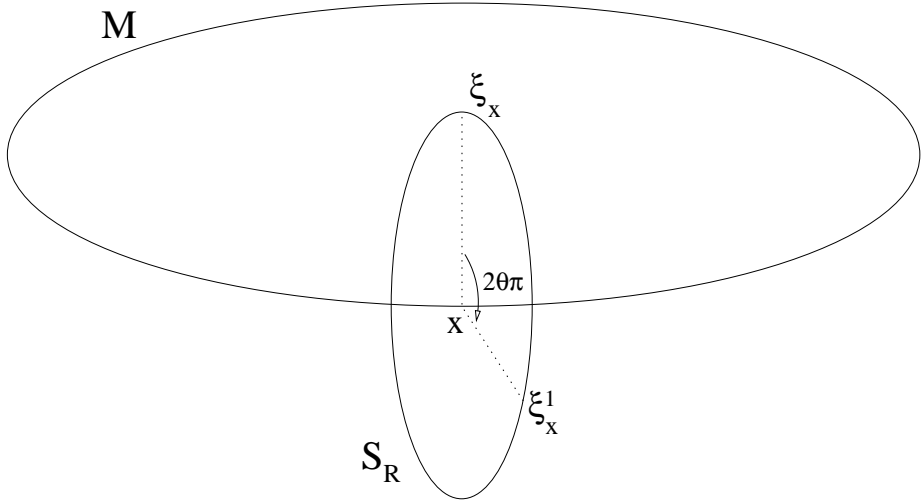}}}}
\end{center}
\caption{ \label{figtore} The $2$-torus $\mathbb{T}_{\xi}$. The
$\xi_x^k$'s form a dense subset of $S_R$.}
\end{figure}

  Consider the trivial
$U(2)$-bundle $P$ over the circle $S^1$ of radius one with fiber
$\cc P^1$, that is to say the set of pure states of
$\aa= \aa_E\ot \aa_I$ with $\aa_E =C^\infty(S^1)$ and $\aa_I = \m2
 $, namely $$\aa=C^\infty(S^1,\m2).$$
Let us equip $P$ with a connection whose associated $1$-form
$A\in\mathfrak{u}(2)$ is constant. For simplicity we restrict to a
matrix $A$ of rank one but the adaptation to a wider class of
connections should be quite straightforward. Once and for all we fix
a basis of $\cc^2$ in which the fundamental representation of $A$ is
written
\begin{equation}
A=\dm{cc} 0&0\\0& -i\theta\fm
\label{connection}
\end{equation}
 where $\theta\in]0,1[$ is a fixed real parameter. Let $[0,2\pi[$ parameterize the circle
 and call $x$ the point with coordinate $0$. Within a
trivialization $(\pi, V)$ the horizontal lift $c$ of the curve
\begin{equation}
\label{cetoile}
 c_*(\tau)=\tau \text{ mod } [2\pi],\quad\; \tau\in]-\infty,+\infty[
 \end{equation}
 with initial condition
 $$V(c(0))=\xi = \dm{c} V_1\\ V_2\fm \in \cc P^1$$
 is the helix
 $c(\tau)=(c_*(\tau), V(\tau))$, where
 \begin{equation}
 \label{vtau}
V(\tau) = \dm{c} V_1 \\ V_2e^{i\theta \tau}\fm.
\end{equation}
 The points of $P$ accessible from $\xi_x = \xi_0 \doteq (\omega_{c_*(0)},
\omega_\xi)$ are the pure states
\begin{equation}
\label{xitau} \xi_\tau\doteq (\omega_{c_*(\tau)},
\omega_{V(\tau)}).
\end{equation}
By the Hopf fibration the fiber $\cc P^1$ is seen to be a two
sphere. Explicitly $\xi$ is the point of $S^2$ with Cartesian
coordinates
\begin{equation}
\label{pourthetazero} x_\xi= 2\re(V_1\overline{V_2}), \; y_\xi=
2\im(V_1\overline{V_2}),\; z_\xi= \abs{V_1}^2 - \abs{V_2}^2 .
\end{equation}
Writing
\begin{equation}
\label{rthetazero}
 2V_1\overline{V_2} \doteq Re^{i\theta_0}
 \end{equation}
  one obtains
$\xi_x$ as the point in the fiber $\pi^{-1}(x)$ with coordinates
$$
x_0 =  R \cos \theta_0,\quad y_0 =  R \sin \theta_0,\quad z_0 =
z_\xi.$$
 The points in the fiber
over $c_*(\tau)$ that are accessible from $\xi_x$ are
\begin{equation}
\label{xitauk}
\xi_\tau^k\doteq \xi_{\tau + 2k\pi},\quad k\in\zz,
\end{equation}
with Hopf
coordinates
\begin{equation}
x_\tau^k \doteq R \cos (\theta_0 - \theta_\tau^k),\quad
y_\tau^k\doteq R \sin (\theta_0 - \theta_\tau^k),\quad
z_\tau^k\doteq z_\xi
\end{equation}
where
$$\theta_\tau^k\doteq \theta(\tau + 2k\pi).$$
All the $\xi_\tau^k$'s are on the
circle $S_R$ of radius $R$ located at the "altitude" $z_\xi$
in $\pi^{-1}(c_*(\tau))$.
Therefore
$$\text{Acc}(\xi_x)\subset \mathbb{T}_{\xi}$$
where
\begin{equation}
\label{txi}
\mathbb{T}_{\xi}\doteq S^1 \times
S_R
\end{equation}
 is the two-dimensional torus (see Figure \ref{figtore}).
Similarly for any $\zeta\in\cc P^1$ such that $z_\xi =
z_\zeta$ one has $\text{Acc}(\zeta_x)\subset \mathbb{T}_{\xi}$. In
fact
\begin{equation}
\label{tz0}
 \mathbb{T}_{\xi} = \underset{\underset{z_\zeta =
z_\xi}{\zeta\in\cc P^1,}}{\bigcup}\text{Acc}(\zeta_x).
\end{equation}
Note that when $\theta$ is irrational $\mathbb{T}_{\xi}$ is the
completion of ${\text{Acc}(\xi_x)}$ with respect to the Euclidean
norm on each $S_R$.

{\prop \label{tori} $\mathbb{T}_{\xi}$ is the connected component
of $\xi_x$ for $d$.}
\\

\noindent{\it Proof.} Let $a_{ij}\in \aa_E$, $i,j=1,2$, be the
components of a selfadjoint element of $\aa$. (\ref{cetoile}) yields
an explicit identification of $\aa_E$ with the algebra of
$2\pi$-periodic complex functions on $\rr$,
\begin{equation}
\label{aij}
 a_{ij}(\tau)\doteq a_{ij}(c_*(\tau)) = a_{ij}(\tau +
2k\pi)\quad k\in\zz \end{equation} with $$a_{ij}(0)= a_{ij}(x).$$
Let dot denote the derivative. Since $M=S^1$ is $1$-dimensional, the
Clifford action reduces to multiplication by $1$ ($\gamma^\mu =
\gamma^1 = 1$) and $[D_E, a_{ij}] = -i\dot{a_{ij}}$. Therefore
\begin{equation}
\label{diraac} i[\dd,a] =  \dm{cc}
\dot{a_{11}}&\dot{a_{12}} + i\theta a_{12}\\
\dot{a_{21}}  - i\theta a_{21}& \dot{a_{22}} \fm
\end{equation}
is zero if and only if $a_{11}=C$, $a_{22}=C'$ are constant and
$a_{12}=\overline{a_{21}}=0$ ($\dot{a_{12}} = -i\theta a_{12}$ has
no other $2\pi$-periodic solution than zero). Under these
conditions
$$\xox(a) = \abs{V_1}^2 C + (1-\abs{V_1}^2)C'$$ differs from
$\yoz(a)$ if and only if $z_\zeta\neq z_\xi$. Hence, identifying
$a_{ij}$ with $\underset{n\rightarrow +\infty}{\lim}(a_n)_{ij}$ in
Lemma \ref{infini}, one obtains that $d(\xox,\yoz)$ is infinite if
and only if $z_\xi\neq z_\zeta$, that is to say
$\yoz\notin\mathbb{T}_{\xi}$. \hfill $\blacksquare$
\\

By the proposition above the connected component $\mathbb{T}_{\xi}$
of $d$ contains, but is distinct from, the connected component
$\text{Acc}(\xi_x)$ of $d_H$. This is enough to establish that the
two metrics are not equal. Furthermore the results of the previous
section strongly suggest that even on $\text{Acc}(\xi_x)$ the two
metrics cannot coincide more than partially. To fix notation let us
consider the distance $d(\xi_x,\xi_\tau)$ with $\xi_\tau\in
\text{Acc}(\xi_x)$ given by (\ref{xitau}) with $\tau > 0$. On the
one hand the function on $\text{Acc}(\xi_x)$
\begin{equation}
\label{L2} L(c(\tau))\doteq d_H(\xox,c(\tau))=\tau
\end{equation}
is not $2\pi$-periodic, hence not in $\aa_E$. Therefore it cannot be used as in
 (\ref{solplate}) to realize the upper bound $d_H$ provided by Proposition \ref{lemme1}. Instead one could be tempted to use the geodesic
distance on $S^1$,
\begin{equation}
\label{dgeosun}
F(\tau)\doteq d_\text{geo}(\xi_x,c_*(\tau)) = \text{min} (\tau \text{ mod }[2\pi],\; (2\pi-\tau)\text{ mod }[2\pi]),
\end{equation}
 but it may help in proving that $d=d_H$ only as long as as $d_H$ equals
$d_\text{geo}$, that is to say as long as $\tau\leq\pi$. Similarly $L
\text{ mod } [2\pi]$ could be efficient till $\tau=2\pi$ but it
has infinite derivative at $2k\pi$ so it cannot be approximated by
some $a_n$ satisfying the commutator norm condition. On the other
hand for fixed $k\in\zz$ the projection of the minimal horizontal
curve between $\xi_\tau^k$ and $\xi_\tau$ is a $K$-fold loop with
$$
K=\left\{
\begin{array}{lcl}
\abs{k}&\text{for}&\theta \text{ irrational}\\
\text{min}\{\abs{k}, \abs{\abs{k}-q}\}&\text{for}&\theta=
\frac{p}{q}
\end{array}
\right.$$ where we assume that $p$ and $q$ are positive relatively
prime with respect to each other and $kp$ is not a multiple of $q$
(otherwise $\xi_\tau^{k}$ coincides with $\xi_\tau$). In any case
when $\abs{k}=1$ then $K=1$ and Proposition \ref{ososip} should not
forbid $d(\xi_\tau, \xi_\tau^{\pm 1})$ from equalling $d_H(\xi_\tau,
\xi_\tau^{\pm 1})=2\pi$. We show below that this is indeed the case
but only when $R=1$. On the contrary as soon as $K
> 3$ Proposition \ref{ososip} certainly forbids $d$ from equalling $d_H$.
In fact the situation is even more restrictive due to the particular
choice (\ref{connection}) of the connection. Since the latter
commutes with the diagonal part $a_1$ of any element $a\in\aa$,
$\xi_\tau^k(a_1) = \xi_{\tau}(a_1)$ for any $k\leq K$. Proposition
\ref{ososip} thus can be written as a system of $K+1$ equations
\begin{eqnarray}
\label{testososipun}
(\xi_\tau^k - \xi_{\tau})(a_o) &=& 2k\pi\\
\label{testososipdeux} \xi_{\tau}(a_o) &=& - \xi_{\tau}(a_1)
\end{eqnarray}
where $a_0= a-a_1$. (\ref{testososipdeux}) simply defines the
diagonal part $a_1$ and one is finally left with $K$ equations
(\ref{testososipun}) constraining the two real components of $a_0$.
Therefore it is most likely that $d$ does not equal $d_H$ as soon as
$K> 2$.
\newline

To make these qualitative suggestions more precise, let us study the
specific example of a "sea-level" (i.e. $z_\xi = 0$) pure state
$\xi$, assuming
\begin{equation}
\abs{V_1} = \abs{V_2} =\frac{1}{\sqrt{2}}.\end{equation} All the
distances on the associated connected component $\mathbb{T}_\xi$ can
be explicitly computed. To do so it is convenient to isolate the
part of the algebra that really enters the game in the computation
of the distances. This is the objective of the following two lemmas.
The first one is of algebraic nature: it deals with our explicit
choice $\aa_I=\m2$ and does not rely on the choice $M=S^1$.

{\lem  \label{lemdiag1} Given $\yoz$ in $\mathbb{T}_{\xi}$,
 the search for the
supremum in the computation of $d(\xox, \yoz)$ can be restricted
to the set of elements
\begin{equation}
\label{diag1} a = f\ii + a_0
\end{equation}
where $\ii$ is the identity of $\m2$, $f\in \aa_E$ vanishes at $x$
and is positive at $y$, while $a_0$ is an element of $\aa$ whose
diagonal terms are both zero and such that
\begin{equation}
\label{xiplus} \yoz(a_0) - \xox(a_0) \geq 0.
\end{equation}}

\noindent{\it Proof.} Let $\;\tilde{ }\;$ denote the operation that
permutes the elements on the diagonal. By
 (\ref{diraac})
\begin{equation}
\label{normdirect} \norm{[\dd,a]}=
\underset{\pm}{\max}\norm{\frac{(\dot{a_{11}} + \dot{a_{22}}) \pm
\sqrt{(\dot{a_{11}}-\dot{a_{22}})^2 + 4\abs{\dot{a_{12}} + i\theta
a_{12}}^2}}{2}}
\end{equation} is invariant under the permutation of $a_{11}$ and
$a_{22}$. Thus $\norm{[\dd,a]} = \norm{[\dd,\tilde{a}]}$ so
\begin{equation}
\label{eqlemm1} \norm{[\dd, \frac{\tilde{a} + a}{2}]}\leq
\norm{[\dd,a]}.
\end{equation}
Meanwhile
\begin{equation}
\xox(\frac{\tilde{a} + a}{2}) = \xox(a) \;\text{ and }\;\quad
\yoz(\frac{\tilde{a} + a}{2}) = \yoz(a)
\end{equation}
therefore the supremum in the distance formula can  be sought on
$$\aa + \tilde{\aa} =C^{\infty}(S^1)\ot\ii + \aa_0$$
where $\aa_0$ is the set of selfadjoint elements of $\aa$ whose
diagonal terms are zero. This fixes eq.(\ref{diag1}). Now if $a=
f\ii + a_0$ attains the supremum then so does $a - f(x)\ii$, hence
the vanishing of $f$ at $x$. Moreover
\begin{eqnarray}
\label{normix}
\norm{[\dd,f\ot\ii]} =
\norm{\dm{cc} 1 &0\\ 0&0\fm [\dd,a]\dm{cc} 1 &0\\ 0&0\fm}&\leq& \norm{[\dd,a]}\\
\label{normix2}
 \norm{[\dd,a_0]}\leq \norm{\dm{cc} 0 &1\\ 0&0\fm [\dd,a]\dm{cc} 0 &1\\ 0&0\fm}&\leq&
 \norm{[\dd,a]},
\end{eqnarray}
so when $a$ satisfies the commutator norm condition so do $f\ii$ and
$a_0$. This implies that $\abs{\xox(a_0) -\yoz(a_0)}$ and
$\abs{\xox(f\ii) -\yoz(f\ii)} =\abs{f(y)}$ are smaller than
\begin{equation}
\label{caraceval} \abs{\xox(a) - \yoz(a)} = \abs{f(y) + \yoz(a_0) -
\xox(a_0)}. \end{equation} In particular, $f(y)$ and $\yoz(a_0) -
\xox(a_0)$ have the same sign, which we assume positive (if not,
consider $-a$ instead of $a$). \hfill $\blacksquare$
\\

\noindent Other simplifications come from the choice of $S^1$ as the
base manifold. Especially the following lemma makes clear the role
played by the functions $L$ and $F$ discussed in
(\ref{L2},\ref{dgeosun}).

{\lem \label{lemdiag2} Let $a= f\ii + a_0$ satisfy the commutator
norm condition, then
\begin{equation}
\label{f} \norm{\dot{f}}\leq 1\;\text{ and }\,  \abs{f(\tau)} \leq
\norm{\dot{f}}F(\tau)
\end{equation}
where $F(\tau)$ is the $2\pi$-periodic function defined on $[0,
2\pi[$ by
\begin{equation}
\label{F}
 F(\tau)\doteq \text{min}\lp \tau, 2\pi - \tau\rp.
\end{equation}
Meanwhile
\begin{equation}
\label{azero} a_0=\dm{cc} 0& ge^{-i\theta L} \\
\overline{g}e^{i\theta L} & 0 \fm
\end{equation}
where $L(\tau)=\tau$ for all $\tau$ in $\rr$
and $g$ is a smooth function on $\rr$ given by
\begin{equation}
\label{g2} g(\tau) = g(0) + \int_{0}^\tau \rho(u)e^{i\phi(u)}du
\end{equation}
with $\rho\in C^\infty(\rr,\rr^+)$, $\norm{\rho}\leq 1$, and
$\phi\in C^\infty(\rr,\rr)$ satisfying
\begin{equation}
\label{krho} \rho(u+2\pi)e^{i\phi(u+2\pi)} =
\rho(u)e^{i(\phi(u)+2\theta\pi)}
\end{equation}
while the integration constant is
\begin{equation}
\label{gperiode} g(0)= \frac{1}{e^{2i\theta\pi} - 1}\int_0^{2\pi}
\rho(u)e^{i\phi(u)} du.
\end{equation}}

\noindent{\it Proof.} (\ref{f}) comes from the
commutator norm condition (\ref{normix}) together with
the $2\pi$-periodicity of $f$ (\ref{aij}), namely
$$f(\tau) = \int_{0}^\tau \dot{f}(u) du = - \int_{\tau}^{2\pi}
\dot{f}(u) du.$$ The explicit form of $a_0$ is obtained by noting
that any complex smooth function $a_{12}\in\aa_E$ can be written
$ge^{-i\theta L}$ where $g\doteq a_{12}e^{i\theta L}\in
C^\infty(\rr)$ satisfies
\begin{equation}
\label{g} g(\tau + 2\pi) = g(\tau)e^{2i\theta\pi}.
\end{equation}
Hence any selfadjoint $a_0$ can be written as in (\ref{azero}),
which yields for the commutator
\begin{equation}
\label{da} [\dd,a_0] = -i\dm{cc}0 & \dot{g}e^{-i\theta L}\\
\dot{\overline{g}}e^{i\theta L} & 0 \fm.
\end{equation}
By (\ref{normix2}) the commutator norm condition implies
$\norm{\dot{g}}\leq 1$, that is to say
\begin{equation}
\label{g2bis}
 g(\tau)= g(0) + \int_{0}^\tau \rho(u)e^{i\phi(u)}du
\end{equation}
where $\rho\in C^{\infty}(\rr,\rr^+)$, $\norm{\rho}\leq 1$, $\phi\in
C^\infty(\rr,\rr)$. The integration constant is fixed by (\ref{g}),
\begin{equation}
\label{glemme} g(0)= \frac{1}{e^{2i\theta\pi} - 1}\lp
\int_{0}^{\tau + 2\pi} \rho(u)e^{i\phi(u)}du - \int_{0}^{\tau}
\rho(u)e^{i(\phi(u) + 2\theta\pi)}du\rp,
\end{equation}
and one extracts (\ref{krho}) from $\frac{d}{d\tau}g(0)=0$. Reinserted in (\ref{glemme})  it
finally yields
(\ref{gperiode}).\hfill $\blacksquare$
\newline

\noindent These lemmas yield the main result of the section: the
computation of all distances on $\mathbb{T}_\xi$.

{\prop \label{s1}
 Let $P$ be the $\cc P^{1}$ trivial bundle over the
circle $S^1$ of radius one with connection (\ref{connection}). Let
 $\xox$ be a point in $P$ at altitude $z_\xi =0$ and $\mathbb{T}_\xi$ its connected component for
the noncommutative geometry distance $d$. For any $\yoz\in
\mathbb{T}_\xi$ there exists an equivalence class of real couples
$(\tau,\theta')\sim (\tau + 2\zz\pi,\theta'- 2\zz\theta \pi)$  such
that
\begin{equation}
\label{zetay}
\yoz =  \dm{cc} 1&0\\
0&e^{i\theta'}\fm\xi_\tau
\end{equation}
where $\xi_\tau$ is given in (\ref{xitau},\ref{vtau}). Without loss
of generality one may assume that $\tau$ is positive (if not,
permute the role played by $\xox$ and $\yoz$) so that
\begin{equation}
\label{tauk} \tau = 2k\pi + \tau_0
\end{equation}
with $k\in\nn$ and $0\leq\tau_0\leq 2\pi$. Then
\begin{equation}
\label{carat}
\displaystyle
d(\xox, \yoz)=
\left\{
\begin{array}{ll}
\max \lp\, X; X + \tau_0 Y\, \rp &\text{when } \tau_0\leq \pi\\
\max \lp\, X; X + (2\pi - \tau_0)Y\,\rp &\text{when } \pi\leq \tau_0
\end{array}
\right.
\end{equation}
in which
\begin{eqnarray}
\label{defx}
X&\doteq& R W_{k+1}
  \tau_0  +
  R W_{k} (2\pi - \tau_0)\\
  \label{defy}
Y&\doteq&1 - RW_{k+1} - RW_k
\end{eqnarray}
with $R$ defined in (\ref{rthetazero}) and
\begin{equation}
\label{wmaxx} W_k  \doteq
\frac{\abs{\sin(k\theta\pi+\frac{\theta'}{2})}}{\abs{\sin\theta\pi}}
\end{equation}
do not depend on the choice of the representantive of the
equivalence class $(\tau,\theta')$.}
\newline

\noindent{\it Proof.} The form (\ref{zetay}) of $\yoz$ comes from
the definition (\ref{tz0}) of $\mathbb{T}_\xi$. It gives, for an
element $a$ of Lemma \ref{lemdiag2},
\begin{equation}
\label{redeux} \abs{\xox(a) -\yoz(a)} = f(\tau) + \Re \lp
Re^{-i\theta_0}(g(\tau)e^{i\theta'} - g(0))\rp
\end{equation}
where we use the definition (\ref{rthetazero}) of $\theta_0$, the
vanishing of $f$ at $x$, the positivity of $f(y)=f(\tau)$ as well as
(\ref{xiplus}). The explicit form (\ref{g2}) of $g$ allows us to
rewrite (\ref{redeux}) as
\begin{equation}
\label{geneux}
 f(\tau) + R \int_0^{\tau}\rho(u)\cos(\phi'(u)) +
 \Re\lp R e^{-i\theta_0} g(0)\lp e^{i\theta'}
-1\rp \rp
\end{equation}
where $$\phi'(u)\doteq \phi(u) - \theta_0 + \theta'.$$ The point
is to find the maximum of (\ref{geneux}) on all the
$2\pi$-periodic $f$ satisfying (\ref{f}), the positive $\rho$,
$\norm{\rho}\leq 1$ and the $\phi$ satisfying (\ref{krho}). To do so we will first find
an upper bound (eqs.
(\ref{boundfinal}) and (\ref{boundfinalbis}) below) and prove that it is the lowest one.

Fixing a pure state $\yoz$ means fixing two values $\theta'$ and
$\tau$ or, equivalently by (\ref{tauk}), fixing $\theta'$, $k$ and
$\tau_0$. The integral term in (\ref{geneux}) then splits into
 \begin{equation}
 \label{gentroisbis}
 \Re  \int_0^{2k\pi} \rho(u)e^{i\phi'(u)}du = \Re \lp \underset{n=0}{\overset{k-1}{\Sigma}}
 e^{2in\theta\pi} \int_0^{2\pi} \rho(u) e^{i\phi'(u)}du\rp
\end{equation}
and
\begin{equation}
\Re  \int_{2k\pi}^{2k\pi + \tau_0} \rho(u)
 e^{i(\phi'(u)})du= \Re \lp e^{2ik\theta\pi}\int_{0}^{\tau_0}
\rho(u) e^{i\phi'(u)}du\rp
 \end{equation}
 that recombine as
\begin{equation}
\label{gentroisdemi}
 S_{k+1}
  \int_{0}^{\tau_0} \rho(u) \cos \phi_{k}(u)\,du + S_{k}
  \int_{\tau_0}^{2\pi} \rho(u) \cos\phi_{k_1}(u)\, du
\end{equation}
where
\begin{equation}
\label{primphik}S_k\doteq \frac{\sin k\theta\pi}{\sin
\theta\pi}\,\text{ and }\, \phi_k(u)\doteq \phi'(u) + k\theta\pi.
\end{equation}
 To compute the real-part term of (\ref{geneux}) one uses the
definition (\ref{gperiode}) of $g(0)$ and obtain
\begin{equation}
\label{gencinqdemi}
S_\frac{1}{2}\int_0^{2\pi} \rho(u)\cos \phi_\frac{1}{2}(u)\, du
\end{equation}
where
$$S_{\frac{1}{2}}\doteq \frac{\sin\theta'/2}{\sin \theta\pi}\, \text{ and }\,\phi_{1/2}(u) \doteq
\phi'(u) - \frac{\theta'}{2} - \theta\pi.$$ (\ref{geneux}) is
rewritten as
\begin{equation}
\label{gentrois}
\abs{\xox(a) - \yoz(a)}=  f(\tau) + R
  \int_{0}^{\tau_0} \rho(u) G_{k+1}(u)\,du  +
  R\int_{\tau_0}^{2\pi} \rho(u)G_{k}(u)\, du
\end{equation}
with
\begin{equation}
\label{gk}
G_k\doteq S_k\cos \phi_{k-1} + S_\frac{1}{2}\cos\phi_\frac{1}{2}.
\end{equation}
The split of the integral makes the search for the lowest upper bound easier.
Calling $W_k$ the maximum of
$\abs{G_k(u)}$ on $[0,2\pi[$, the positivity of $\rho$ makes (\ref{gentrois}) bounded by
\begin{equation}
\label{genquatre}
f(\tau) + R W_{k+1}
  \int_{0}^{\tau_0} \rho(u) \,du  +
  R W_{k}\int_{\tau_0}^{2\pi} \rho(u)\, du.
\end{equation}
Now (\ref{normdirect}) with $a_{11} = a_{22} = f$ and
$\abs{\dot{a_{21}}+ i\theta a_{12}}=\rho$ yields
\begin{eqnarray}
\abs{\dot{f}(u) + \rho(u)}\leq 1 &\text{whenever}& \abs{\dot{f}(u)}\geq 0\\
\abs{\dot{f}(u) - \rho(u)}\leq 1 &\text{whenever}&\abs{\dot{f}(u)}\leq 0
\end{eqnarray}
for any $u\in\rr$, that is to say
 \begin{equation}
 \label{frho}
\rho \leq 1 - \abs{\dot{f}}.
\end{equation}
Therefore
\begin{equation}
\label{testzero}
\int_{0}^{\tau_0} \rho(u)\, du \leq \tau_0 - \int_{0}^{\tau_0} \abs{\dot{f}}.
\end{equation}
Moreover  $f(\tau)=f(\tau_0)$ ($2\pi$-periodicity of $f$) is
positive by Lemma \ref{lemdiag1} so
\begin{equation}
\label{fpositive} -f(\tau_0) = -\abs{f(\tau_0)} \geq
-\int_{0}^{\tau_0} \abs{\dot{f}(u)}\, du.
\end{equation}
Hence (\ref{testzero}) gives
\begin{equation}
\label{testun}
\int_{0}^{\tau_0} \rho(u)\, du \leq \tau_0 -f(\tau_0).
\end{equation}
Similarly
$$
\int_{\tau_0}^{2\pi}\abs{\dot{f}(u)} du \geq \abs{
\int_{\tau_0}^{2\pi}\dot{f}(u) du} = \abs{-
\int_{0}^{\tau_0}\dot{f}(u)\, du}= f(\tau_0)
$$
hence
\begin{equation}
\label{testdeux}
\int_{\tau_0}^{2\pi} \rho(u)\, du \leq 2\pi - \tau_0 - f(\tau_0).
\end{equation}
Back to (\ref{genquatre}), equations (\ref{testun}) and
(\ref{testdeux}) yield the bound
\begin{equation}
\label{gensept}
f(\tau_0) Y + X
\end{equation}
where $X$ is defined in (\ref{defx}) and $Y$ in (\ref{defy}).
By (\ref{f}) and in case
\begin{equation}
\label{condfinale}
Y \geq 0,
\end{equation}
(\ref{gensept}) yields
\begin{equation}
\label{boundfinal}
\abs{\xox(a) - \yoz(a)} \leq \left\{
\begin{array}{ll}
X + \tau_0 Y   &\text{for } 0\leq \tau_0\leq \pi\\
X + (2\pi-\tau_0)Y &\text{for } \pi\leq \tau_0\leq 2\pi\end{array}
\right. .
\end{equation}
When $Y\leq 0$,
\begin{equation}
\label{boundfinalbis} \abs{\xox(a) - \yoz(a)} \leq X.
\end{equation}

These are the announced lowest upper bounds. To convince ourselves
let us build a sequence $a_n$ that realizes (\ref{boundfinal}) or
(\ref{boundfinalbis}) at the limit $n\rightarrow +\infty$. As a
preliminary step note that an easy calculation from (\ref{gk})
yields
$$G_k = A_k \cos \phi' + B_k\sin \phi'$$
where
\begin{eqnarray}
A_k&\doteq& S_\frac{1}{2}\cos(\frac{\theta'}{2}+\theta\pi) + S_k\cos(k-1)\theta\pi\\
B_k&\doteq& S_\frac{1}{2}\sin(\frac{\theta'}{2}+\theta\pi) - S_k\sin(k-1)\theta\pi.
\end{eqnarray}
$G_k$ attains its maximum value
\begin{equation}
\label{wmax}
W_k  \doteq \abs{A_k}\sqrt{1 + \frac{\abs{B_k}^2}{\abs{A_k}^2}} =
\frac{\abs{\sin(k\theta\pi+\frac{\theta'}{2})}}{\abs{\sin(\theta\pi)}}
\end{equation}
when{\footnote{The ambiguity in the explicit form of $\Phi_k$ is not relevant. Depending
on the respective signs of $A_k$ and $B_k$, one choice yields $W_k$ whereas the other one yields
$-W_K$. What is important is the existence of a well defined value $\Phi_k$
such that $A_k\cos \Phi_k + B_k\sin\Phi_k = W_k$.}}
\begin{equation}
\label{phimax} \phi'= \Phi_k \doteq \text{Arctan}\frac{B_k}{A_k}\,
\text{ or } \, \text{Arctan}\frac{B_k}{A_k}+\pi.
\end{equation}
Let then
$$a_n = \dm{cc} f_n & g_n e^{-i\theta L}\\ \overline{g}_n e^{i\theta L}&f_n\fm$$
be a sequence of elements of $\aa$ that depend on the fixed value
$\tau = 2k\pi + \tau_0$ in the following way:
 in case (\ref{condfinale}) is fulfilled and $\tau_0\leq \pi$, $f_n$ approximates from below  the $2\pi$-periodic
function
\begin{equation}
\label{fmoins} f_{-} (t) = \left\{\begin{array}{ll}
t&\text{ for }\, 0\leq t\leq \tau_0 \\
\tau_0 - C(t-\tau_0)&\text{ for }\, \tau_0\leq t\leq 2\pi
\end{array}\right.
\end{equation}
with $$C\doteq\frac{\tau _0}{2\pi -\tau_0}.$$
In case  $\tau_0\geq \pi$, $f_n$ approximates
\begin{equation}
\label{fplus} f_{+} (t) = \left\{\begin{array}{ll}
\frac{t}{C}&\text{ for }\, 0\leq t\leq \tau_0 \\
2\pi- t &\text{ for }\, \tau_0\leq t\leq 2\pi
\end{array}\right. .
\end{equation}
\begin{figure}[h]
\begin{center}
\mbox{\rotatebox{0}{\scalebox{.7}{\includegraphics{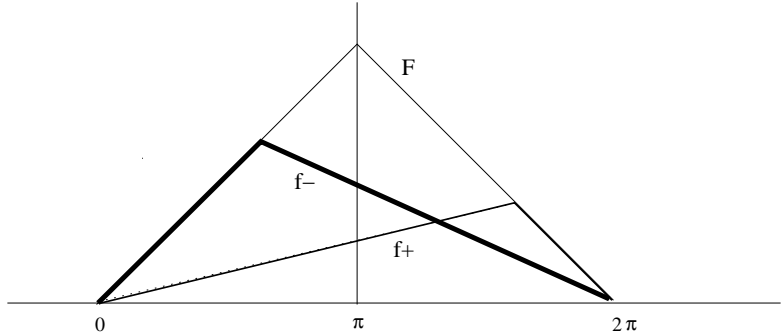}}}}
\end{center}
\caption{\label{3f}The functions $F$, $f_+$ and $f_-$.}
\end{figure}

\noindent When (\ref{condfinale}) is not fulfilled, $f_n = f_0$ is
simply the zero function. In any case and whatever $\tau_0$, $g_n$
is defined via (\ref{g2}) and (\ref{gperiode}), replacing $\phi$
with a sequence $\phi_n$ approximating the step function $\Phi$ of
width $2\pi$ and height $2\theta\pi$ defined on $[0,2\pi[$ by
\begin{equation}
\label{bisphi} \Phi(u) =\left\{ \begin{array}{ll} \Phi_{k+1} +
\theta_0 - \theta' 
&\text{ for } 0 \leq u < \tau_0\\
\Phi_{k} + \theta_0 - \theta' 
&\text{ for } \tau_0 < u <\pi 
\end{array} \right. ,
\end{equation}
and replacing $\rho$ with a sequence $\rho_n$ approximating the
$2\pi$-periodic function
\begin{equation}
\label{rrho} \Gamma_I = 1 -\abs{\dot{f_I}}
\end{equation}
where $I=+, -$ or $0$. By construction the $a_n$'s satisfy the
commutator norm condition. In particular the fact that
$\underset{n\rightarrow +\infty}{\text{lim}}\rho_n$ and $\Phi$ are
step functions
 is not problematic since their derivatives are not constrained by the commutator.
 For technical details on how to approximate step functions
by sequences of smooth functions, the reader is invited to consult
classical textbooks such as [\citelow{doubro}]. The last point is to
check that
\begin{equation}
\underset{n\rightarrow +\infty}{\text{lim}} \abs{\xox(a_n) - \yoz(a_n)} = (\ref{boundfinal}).
\end{equation}
This is a simple notation exercise: (\ref{primphik}) gives
\begin{eqnarray}
\label{phi}
\phi'(u)&=&\Phi_{k+1}\;\text{ for }\, 0 \leq u < \tau_0\\
\phi'(u)  &=&\Phi_{k}\quad\text{ for }\, \tau_0 < u < 2\pi.
\end{eqnarray}
Therefore, by (\ref{gentrois}) together with (\ref{rrho}),
\begin{eqnarray}
\nonumber
\underset{n\rightarrow +\infty}{\text{lim}} \abs{\xox(a_n) - \yoz(a_n)}&=&
  f_I(\tau_0) + RW_{k+1}
  \int_{0}^{\tau_0} \Gamma_I(u) \,du  +
  RW_k\int_{\tau_0}^{2\pi} \Gamma_I(u)\, du\\
  \nonumber
 &= &f_I(\tau_0) -  RW_{k+1}\int_{0}^{\tau_0}\abs{\dot{f_\pm}}\,du -
 RW_k\int_{\tau_0}^{2\pi} \abs{\dot{f_I}}\,du \\
 \label{checkmoins}& & + RW_{k+1}
  \tau_0  +  RW_k(2\pi-\tau_0).
\end{eqnarray}
When $(\ref{condfinale})$ is fulfilled and $\tau_0\leq\pi$, the
subscript of $f$ is minus and (\ref{fmoins}) makes
(\ref{checkmoins}) equal to
$$\tau_0 -  RW_{k+1}\tau_0 -
 RW_k(2\pi - \tau_0) C + RW_{k+1}
  \tau_0  +  RW_k(2\pi-\tau_0)
$$
which is exactly the first line of (\ref{boundfinal}). Similarly for
$\tau_0\geq\pi$, the subscript turns to $+$ and (\ref{fplus}) yields
for (\ref{checkmoins})
$$(2\pi - \tau_0)  -  RW_{k+1}\frac{\tau_0}{C} -
 RW_k(2\pi - \tau_0) + RW_{k+1}
  \tau_0  +  RW_k(2\pi-\tau_0),
$$
which is nothing but the second line of (\ref{boundfinal}). Finally,
when (\ref{condfinale}) is not fulfilled, $f_I=\dot{f}_I=0$ and
(\ref{checkmoins}) equals (\ref{boundfinalbis}). \hfill
$\blacksquare$
\\

 Let us check the coherence of our result by noticing
that for $\tau_0= \pi$ both formulas of (\ref{carat}) agree and
yield
{\che
$d(\xox, \zeta_\pi) = \max\lp X, X+\pi Y\rp=\max\lp \pi R(W_{k+1} + W_k);\, \pi\rp.$}
\\

\noindent Similarly for a given $k$ and
$\tau_0=2\pi$, the second line of (\ref{carat}) agrees with the first line
with $k+1$ and $\tau_0=0$, namely
{\che
\label{circle}
$d(\xox, \zeta_{2k\pi + 2\pi}) = 2\pi R W_{k+1} =
d(\xox, \zeta_{2(k+1)\pi + 0}).$}\\

\noindent This is nothing but the restriction of $d$ to the fiber
over $x$. Its extreme simplicity (no "max" is involved) indicates
that the noncommutative metric is better understood fiberwise. We
shall see in the next section that this is the main difference from
the Carnot-Carath\'eodory metric. Another check, and certainly the
best guarantee that Proposition \ref{s1} is true, is to directly
verify that formula (\ref{carat}) does define a metric: the
vanishing of $d$ when $\yoz = \xox$ is obvious; the invariance under
the exchange $\xox \longleftrightarrow \yoz$ is not testable since
the symmetry $\tau\longleftrightarrow -\tau$ is broken from the
beginning by the specification that $\tau$ is positive. There
remains the triangle inequality.

{\che For any $\zeta_1,\zeta_2\in \mathbb{T}_\xi$, $d(\xi_x,\zeta_2)
\leq d(\xi_x,\zeta_1) + d(\zeta_1,\zeta_2).$}
\\

\noindent{\it Proof.} Let $\zeta_{\tau_i}$,  $i=1,2$, be two pure states defined by $\tau_i = 2\pi
k_i + t_i$ and $\theta'_i$, labeled in such a way that $\tau_1\leq\tau_2$. The point is to check
that
\begin{equation}
\label{ddelta}
\Delta\doteq d(\xox,\zeta_{\tau_1}) + d(\zeta_{\tau_1},\zeta_{\tau_2}) - d(\xox,\zeta_{\tau_2})
\end{equation}
is positive. Proposition \ref{s1} is invariant under translation
(i.e. a reparameterization of the circle $\tau\rightarrow \tau +
\text{constant}$), which means that
$d(\zeta_{\tau_1},\zeta_{\tau_2})$ is given by formula (\ref{carat})
with $W_k$ replaced by
$$W_{k_{12}} \doteq
\frac{\abs{\sin(k_{12}\theta\pi+\frac{\theta'_2 - \theta'_1}{2})}}{\abs{\sin\theta\pi}}
$$
and $\tau_0$ replaced by $t_{12}$. Here $k_{12}$ and $t_{12}$ are such that
$\tau_{12}\doteq \tau_2 - \tau_1 \doteq  2k_{12}\pi + t_{12}$. Explicitly
\begin{eqnarray}
\label{t121}
k_{12} = k_2 - k_1,&\; t_{12} = t_2 - t_1 &\text{ if }\, t_1\leq t_2\\
\label{t122}
k_{12} = k_2 - k_1 -1,&\; t_{12} = 2\pi + t_2 - t_1 &\text{ if }\, t_2\leq t_1.
\end{eqnarray}
Let $X_i, Y_i$, $i\in\{1,2,12\},$ denote (\ref{defx}) and (\ref{defy})
in which $k$ is replaced by $k_i$. The only difficulty in checking that (\ref{ddelta}) is
positive is the quite large number of possible expressions for
$\Delta$: one for each combination of the signs of the $Y_i$'s and $t_i -\pi$. A simple way to reduce
the number of cases under investigation is to decorate $\Delta$ with three arrows indicating
whether $Y_1$, $Y_{12}$ and $Y_2$ respectively are positive (upper arrow) or negative
(lower arrow). For instance
$\Delta_{\uparrow\uparrow\downarrow}$ denotes the value of $\Delta$ when $Y_1\geq 0$, $Y_{12}\geq 0$,
and $Y_2\leq 0$. Let us also use $\tilde{\Delta}$ decorated with arrows to denote the formal expression (\ref{ddelta}) in which
$d(\xox,\xi_{\tau_1})$, $d(\zeta_{\tau_1},\zeta_{\tau_2})$ and
$d(\xox,\zeta_{\tau_2})$ are replaced either by $X_i + t_i^m Y_i$ (upper arrow) or by $X_i$
(lower arrow). Here $t_i^m \doteq \min\lp t_i, 2\pi - t_i\rp$. For instance
\begin{eqnarray}
\tilde{\Delta}_{\uparrow\uparrow\uparrow}&=& \tilde{\Delta}_{\downarrow\uparrow\uparrow} + t_1^m Y_1\\
&=& \tilde{\Delta}_{\uparrow\downarrow\uparrow} + t_2^m Y_2\\
&=& \tilde{\Delta}_{\downarrow\downarrow\uparrow} + t_1^m Y_1 + t_2^m Y_2.
\end{eqnarray}
Now suppose that $Y_1$, $Y_{12}$, $Y_2$ are all positive,
then
$$\Delta = \Delta_{\uparrow\uparrow\uparrow}=\tilde{\Delta}_{\uparrow\uparrow\uparrow}\geq
\left\{\begin{array}{c} \tilde{\Delta}_{\downarrow\uparrow\uparrow}\\
 \tilde{\Delta}_{\uparrow\downarrow\uparrow}\\\tilde{\Delta}_{\downarrow\downarrow\uparrow}
 \end{array}\right. .
$$
Changing the sign of $Y_1\leq 0$ and $Y_{12}$ yields
$$\Delta = \Delta_{\downarrow\downarrow\uparrow}=\tilde{\Delta}_{\downarrow\downarrow\uparrow}\geq
\left\{\begin{array}{c} \tilde{\Delta}_{\uparrow\downarrow\uparrow}\\
 \tilde{\Delta}_{\downarrow\uparrow\uparrow}\\\tilde{\Delta}_{\uparrow\uparrow\uparrow}
 \end{array}\right. .
$$
Therefore, if one is able to show {\it without using the sign of
$Y_1$ or the sign of $Y_{12}$} that
$\tilde{\Delta}_{\uparrow\uparrow\uparrow}$ is positive, one proves
that both $\Delta_{\uparrow\uparrow\uparrow}$ and
$\Delta_{\downarrow\downarrow\uparrow}$ are positive. In fact
showing that one of the
$\tilde{\Delta}_{\updownarrow\updownarrow\uparrow}$'s is positive is
enough to prove that all the
$\Delta_{\updownarrow\updownarrow\uparrow}$'s are positive (here
$\updownarrow$ means either $\uparrow$ or $\downarrow$). Of course
the same is true with
$\tilde{\Delta}_{\updownarrow\updownarrow\downarrow}$ so that, at
the end, one just has to check the inequality of the triangle for
one of the $\tilde{\Delta}_{\updownarrow\updownarrow\uparrow}$ and
one of the $\tilde{\Delta}_{\updownarrow\updownarrow\downarrow}$.

Let us begin by
$\tilde{\Delta}_{\updownarrow\updownarrow\downarrow}$, assuming
first $t_1\leq t_2$. With $W_i\doteq W_{k_i}$, $W_{i+1}\doteq W_{k_i
+1}$, (\ref{t121}) yields
\begin{eqnarray*}
R^{-1}\tilde{\Delta}_{\downarrow\downarrow\downarrow}
&=&W_{1+1}t_1 + W_{1}(2\pi - t_1) + W_{12+1}t_{12} + W_{12}(2\pi - t_{12})\\
& &- W_{2+1}t_2 -  W_{2}(2\pi - t_2)\\
&=& (2\pi - t_2)(W_{1}+ W_{{12}} - W_{2}) + t_{12}(W_{1} + W_{{12}+1}
- W_{2+1})\\ & &+ t_1(W_{1+1} + W_{{12}} - W_{2+1})
\end{eqnarray*}
which is positive since
{\footnote{this comes from $\abs{\sin(a+b)}\leq\abs{\sin a} + \abs{\sin b}$ with $a=(k_2-k_1)\theta\pi + \theta'_2-\theta'_1$
and $b=k_1\theta\pi + \theta'_1$}}
\begin{equation}\label{3W}
{W_{k_2}} \leq {W_{k_1}} + {W_{k_{12}}}
\end{equation}
and similar equations for the other indices. Assuming now $t_2\leq t_1$, (\ref{t122}) yields
\begin{eqnarray*}
R^{-1}\tilde{\Delta}_{\downarrow\downarrow\downarrow}
&=&t_2(W_{1+1} + W_{12+1} - W_{2+1}) + (2\pi - t_1)(W_1 + W_{12+1} - W_2)\\
&&+ (2\pi - t_{12})(W_{12}+ W_{1+1} - W_2)
\end{eqnarray*}
which is also positive by equations similar to (\ref{3W}) (be careful to use the definition
(\ref{t122}) of $k_{12}$ and no longer definition (\ref{t121})). Thus, whatever $t_1$ and $t_2$,
$\tilde{\Delta}_{\updownarrow\updownarrow\downarrow}$ is positive and the triangle inequality is checked for all
the configurations $\updownarrow\updownarrow\downarrow$ of the $Y_i$'s.

Things are slightly more complicated for the configurations $\updownarrow\updownarrow\uparrow$ for
one also
has to deal with the signs of $t_i - \pi$. First assume $t_1\leq t_2$:
\begin{itemize}
\item $\ t_1\leq t_2\leq\pi$ (implies $t_{12}\leq\pi$),
 \begin{eqnarray*}
         (2R)^{-1}\tilde{\Delta}_{\uparrow\uparrow\uparrow}
        &=& {W_1}(\pi -t_1) + {W_{{12}}}(\pi -t_{12})  -
        {W_{2}}(\pi -t_{2})\\
        &\geq& (\pi - t_2)( {W_{1}} + {W_{{12}}} - W_{2}).
        \end{eqnarray*}

\item $\pi \leq\ t_1\leq t_2$ (implies $t_{12}\leq\pi$),
  \begin{eqnarray*}
        \tilde{\Delta}_{\uparrow\uparrow\uparrow}
        &=& 2R(W_{1+1}(t_1-\pi)  + W_{{12}}(\pi - t_{12})  - W_{2+1}(t_2-\pi)) + 2(t_2 - t_1)
        \\
        &\geq&
        2R(t_1-\pi)(W_{1+1}  + W_{{12}}- W_{2+1}) + 2(t_2 - t_1)(1 -
        RW_{2+1}).
        \end{eqnarray*}
    \item $t_1\leq\pi\leq t_2$ and  $t_{12}\leq\pi$,
        $$
       \tilde{\Delta}_{\uparrow\uparrow\uparrow}
        = 2R(W_{1}(\pi -t_1)  + W_{{12}}(\pi - t_{12})) +2(t_2 -\pi)(1 - RW_{2+1}).
        $$
        \item $t_1\leq\pi\leq t_2$ and  $t_{12}\geq\pi$,
        \begin{eqnarray*}
        \tilde{\Delta}_{\uparrow\uparrow\uparrow}
        &=& 2R( W_{1}(\pi-t_1) + W_{{12}+1}(t_{12} - \pi)
        - W_{{2}+1}(t_{12} - \pi))  + t_1(1 - 2RW_{{2}+1})\\
        &\geq& 2R(t_{12}-\pi)(W_{1} + W_{{12}+1} - W_{{2}+1})  + 2t_1(1 - RW_{{2}+1}).
        \end{eqnarray*}
 \end{itemize}
These five expressions are positive by (\ref{3W}) and the positivity
of $Y_2$. Similarly, in case $t_2\leq t_1$:
    \begin{itemize}
   \item $t_2\leq t_1 \leq \pi$ (implies $t_{12}\geq\pi$),
        \begin{eqnarray*}
        \tilde{\Delta}_{\uparrow\uparrow\uparrow}
        &=& 2R(W_{1}(\pi -t_1) + W_{{12}+1}(t_{12}-\pi) -
        W_{2}(\pi-t_1)) + 2(t_1 - t_2)(1 - R{W_{2}})\\
        &\geq& 2R(\pi -t_1)\lp W_{1} + W_{{12}+1} -
        W_{2}\rp + 2(t_1 - t_2)\lp 1 - R{W_{2}}\rp.
        \end{eqnarray*}
    \item $\pi \leq\ t_2\leq t_1$ (implies $t_{12}\geq\pi$),
        \begin{eqnarray*}
        (2R)^{-1}\tilde{\Delta}_{\uparrow\uparrow\uparrow}
        &=& W_{1+1}(t_1-\pi) + W_{{12}+1}(t_{12} - \pi)
        - W_{2+1}(t_2-\pi)
        \\
        &\geq&
        (t_2-\pi) (W_{1+1} + W_{{12}+1}
        - W_{2+1}).
        \end{eqnarray*}
    \item $t_2\leq\pi\leq t_1$ and  $t_{12}\leq\pi$,
    \begin{eqnarray*}
        \tilde{\Delta}_{\uparrow\uparrow\uparrow}
        &=& 2R(W_{1+1}(t_1 - \pi) + W_{{12}}(\pi - t_{12}) -W_2(\pi -t_{12}))
        + 2(2\pi - t_1)(1-RW_2)\\
        &\geq& 2R(\pi - t_{12})( W_{1+1} + W_{{12}} -W_2)
        + 2(2\pi - t_1)(1-RW_2).
        \end{eqnarray*}
 \item $t_2\leq\pi\leq t_1$ and  $t_{12}\geq\pi$,
        \begin{eqnarray*}
        \tilde{\Delta}_{\uparrow\uparrow\uparrow}
        &=& 2RW_{1+1}(t_1 - \pi) + 2RW_{{12}+1}(t_{12}-\pi)
          + 2(\pi  - t_2)(1 - RW_{2}).
        \end{eqnarray*}\hfill $\blacksquare$
\end{itemize}

\noindent The proof above is long but we believe it is important to
convince oneself that formula \ref{s1} does define a metric, which
is not obvious at first sight. As a final test, let us come back to
the beginning of this section and verify Lemma \ref{lemme1}. {\che
$d(\xox, \yoz) \leq d_H(\xox, \yoz)\, \text{ for any }\,
\yoz\in\text{Acc}(\xox).$}
\\

\noindent{\it Proof.}
Let $\yoz = \xi_\tau$. Then $d_H(\xox,\xi_\tau) = 2k\pi + \tau_0$ so that
\begin{equation}
\label{testddh}
d(\xox, \xi_\tau)-d_H(\xox,\xi_\tau) \left\{\begin{array}{lll}
=&2RW_k(\pi-\tau_0) - 2k\pi& \\
&\leq 2\pi(W_k - k) &\text{
when } Y\geq 0,\, \tau_0\leq\pi,\\
=& 2RW_{k+1}(\tau_0 -\pi) - 2(\tau_0 -\pi) - 2k\pi&\\
&\leq -2(\tau_0 -\pi)W_{k} - 2k\pi &\text{ when } Y\geq 0,\, \tau_0\geq\pi,\\
=& \tau_0(RW_{k+1} - RW_{k} -1)\\
&+ 2\pi(RW_{k} -k)& \text{ when } Y\leq 0.
\end{array}\right .
\end{equation}
These three expressions are negative by (\ref{3W}) and
{\footnote{obvious for $k\leq 1$, then by induction} $\abs{\sin
k\theta\pi}\leq k\abs{\sin \theta\pi}.$\hfill $\blacksquare$

\begin{figure}[ht]
\begin{center}
\mbox{\rotatebox{0}{\scalebox{1}{\includegraphics{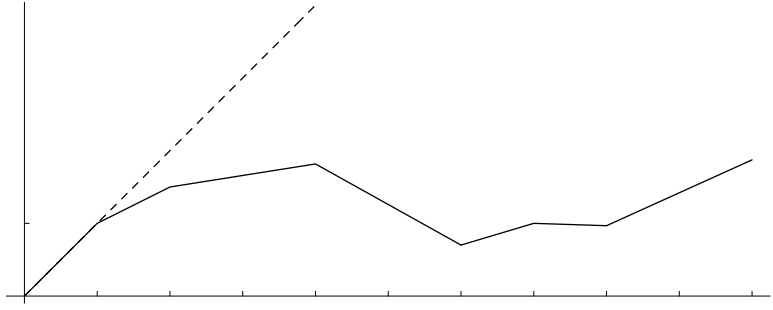}}}}
\end{center}
\caption{\label{dgnc}$d(\xox,\xi_\tau)\,$ as a function of $\tau$
for $\theta = \frac{1}{\sqrt{2}}$, $R=0.75$. The dashed line is
$d_H(\xox,\xi_\tau)$. The unit is $\pi$.}
\end{figure}

\section{\label{interpretation}Interpretation: a smooth cardio-torus \label{torecoeur}}

This section aims at analyzing the result of Proposition \ref{s1}.
We first compare $d$ to $d_H$ on $\text{Acc}(\xox)$ (corollaries
\ref{dense} and \ref{erreur}), then study the restriction of $d$ to
the fiber over $x$ and to the base $M=S^1$. The reader may wonder
why we do not systematically replace $R$ by its value $1$. The point
is that for two states on the same fiber ($y=x$) the diagonal part
of $a$ does not play any role so that Proposition \ref{s1} is valid
also for non vanishing $z_\xi$. Also, for $y\neq x$ some
calculations\cite{caratun} show that \ref{s1} is still valid for
non-zero $z_\xi$ as long as $2V^2_i - R(W_{k+1} + W_k)$ is positive
for both $i=1,2$. This is the reason why, in the following
discussion, we keep writing $R$.

\subsection{The shape of $\mathbb{T}_\xi$}

Taking $\yoz$ in $\text{Acc}(\xox)$
amounts to setting $\theta'=0$. $W_k$ is replaced by
$$S_k \doteq \frac{\abs{\sin k\theta\pi}}{\abs{\sin \theta\pi}}$$
and proposition \ref{s1} is rewritten in a somehow more readable
fashion.

{\cor \label{dense} Let $\yoz = \xi_\tau\in \text{Acc}(\xox)$, with $\tau=2k\pi+\tau_0.$
For $k$ such that $S_{k+1} + S_k\leq \frac{1}{R}$,
\begin{equation}
\label{caratore}
\displaystyle
d(\xox, \xi_\tau)=   \left\{
\begin{array}{ll}\displaystyle
2R S_{k}(\pi -\tau_0)  + \tau_0     &\text{when } \tau_0\leq \pi\\
2R S_{k+1}(\tau_0 -\pi) +  2\pi - \tau_0&\text{when } \pi\leq \tau_0\end{array}
\right. .
\end{equation}
For $k$ such that $S_{k+1} + S_k \geq \frac{1}{R}$,
$$
\displaystyle
d(\xox, \xi_\tau)=   RS_{k+1}\tau_0 + RS_k(2\pi - \tau_0).$$}

\noindent It is easy to see on which part of $P$ the noncommutative geometry metric
and the Carnot-Carath\'eodory one coincide.

{\cor \label{erreur} For any $R$, $d(\xox, \xi_\tau) = d_H(\xox,
\xi_\tau)$ for $\tau\in[0,\pi]$. Moreover if $R=1$ the two metrics
are also equal for $\tau\in[\pi,2\pi]$. These are the only
situations in which $d=d_H$.}
\\

\noindent {\it Proof.} $S_0 = 0$, $S_1 =1$ and by construction $R\leq 1$. Therefore for $k=0$,
$S_{k+1} + S_k = 1 \leq \frac{1}{R}$ so
\begin{equation}
\displaystyle
d(\xox, \xi_\tau)=   \left\{
\begin{array}{ll}\displaystyle
\tau_0 = d_H(\xox,\xi_{\tau_0})    &\text{when } \tau_0\leq \pi\\
2\pi(1-R) + \tau_0(2R -1) &\text{when } \pi\leq \tau_0\end{array}
\right.
\end{equation}
which yields the equality of $d$ and $d_H$ for the indicated values of $\tau$ and $R$.
From check 4 in the preceding section, $d$ may equal $d_H$ only if
$S_k=k$, i.e. $k=0$ or $1$. When $k=1$, $S_k + S_{k+1}\geq 1$ and the last line of (\ref{testddh})
gives the difference $\delta$ between $d$ and $d_H$,
\begin{equation}
\label{raslebol}
\delta = \tau_0(RS_2 - R -1)+ 2\pi(R-1).
\end{equation}
$S_2\leq 2$  so $\delta \leq (R -1)(2\pi + \tau)$. $\delta$ may
vanish only if $R=1$ and, going back to (\ref{raslebol}), only if
$\tau_0=0$. \hfill$\blacksquare$
\\

This result is more restrictive that what was expected from
Proposition \ref{ososip} revisited in (\ref{testososipun}), namely
that $d$ may equal $d_H$ as long as $c$ does not have sequences of
more than $2$ self-intersecting points, i.e. up to
$\tau_{\text{max}}= 4\pi + \tau_0$. It seems that Proposition
\ref{ososip} alone is not sufficient to show that
$\tau_{\text{max}}\leq 2\pi$. At best one can obtain
\begin{equation}
\label{taumax}
\tau_{\text{max}}< 4\pi.
\end{equation}
Although (\ref{taumax}) is not in se an interesting result but
simply a weaker formulation of Corollary \ref{erreur}, we believe it
is interesting to see how far Proposition \ref{ososip} can lead.
This could be the starting point for a generalization of the results
of this paper to manifolds other than $S^1$. Let $G, \overline{G}$
be the off-diagonal components of $a$. (\ref{testososipun}) is
rewritten as
\begin{equation}
\label{testososip}
 \Im\lp G(\tau) e^{i(\theta\tau-\theta_0)}e^{ik\theta\pi}\rp= -\frac{k\pi}{R\sin k\theta\pi}
 \end{equation}
for any $k=1,...,K$. For $K=2$ this system has a unique solution
\begin{equation}
\label{solososip}
G(\tau) = Ce^{-i\theta\tau}e^{i(\theta_0 - \frac{\pi}{2})}
\end{equation}
where
$$C \doteq -\frac{2\pi }{R\sin 2\theta\pi}
$$
is a constant. Therefore $\xi_\tau(a_0) = \Re\lp
e^{i(\theta\tau-\theta_0)}G(\tau)\rp = 0$ so that, by
(\ref{testososipdeux}), $\xi_\tau(a) =0$. By Proposition
\ref{ososip} this is possible only for $\tau=0$. Hence there cannot
be more than one sequence of 2 self-intersecting points, hence
(\ref{taumax}).

In any case, when $\tau$ is greater than $2\pi$, $d$ strongly
differs from $d_H$. While the latter is unbounded, the former is
bounded,
$$d(\xox,\yoz)\leq\text{max}(\frac{2\pi R}{\abs{\sin\theta \pi}},\pi).$$
As illustrated in figure \ref{dgnc}, $\text{Acc}(\xox)$ viewed as a $1$-dimensional
object looks like a straight line when it is equipped with $d_H$,  whereas it looks rather
chaotic when it is equipped with $d$.

\subsection{The shape of the fiber}

From a fiberwise point of view the situation drastically changes.
Parameterizing the fiber $S_x$ over $x$ by
$$\phi \doteq 2k\theta\pi  + \theta' \text{ mod }[2\pi],$$
one obtains a very simple expression for the noncommutative distance,
\begin{equation}
\label{dphi}
d(0,\phi) =\frac{2\pi R}{\abs{\sin \theta\pi}} \sin \frac {\phi}{2}.
\end{equation}
For those points of $S_x$ which are accessible from $\xox$, namely
for $\theta' = 0$, the Carnot-Carath\'eodory metric is
$$d_H(0,\phi) = 2k\pi.
$$
Hence, when $\theta$ is irrational and in any neighborhood of $\xox
= 0$ in the Euclidean topology of $S_x$, it is always possible to
find some
$$\phi_k\doteq \xi_0^k = 2k\theta\pi \text{ mod }[2\pi]$$ which are arbitrarily Carnot-Carath\'eodory-far
from $\xox$. In other terms $d_H$ destroys the $S^1$ structure of
the fiber. On the contrary $d$ keeps it in mind in a rather
intriguing way. Let us compare $d$ to the Euclidean distance $d_E$
on the circle of radius
\begin{equation}
\label{R}
\mathcal{R}\doteq \frac{2R}{\abs{\sin \theta\pi}}.
\end{equation}
 At the cut-locus $\phi=\pi$, the two distances are equal
but whereas $d_E(0,.)$ is not smooth, the noncommutative geometry
distance {\it is} smooth (cf Figure \ref{ds}).
\newline

\begin{figure}[hh]
\hspace{-.5cm}
\mbox{\rotatebox{0}{\scalebox{.8}{\includegraphics{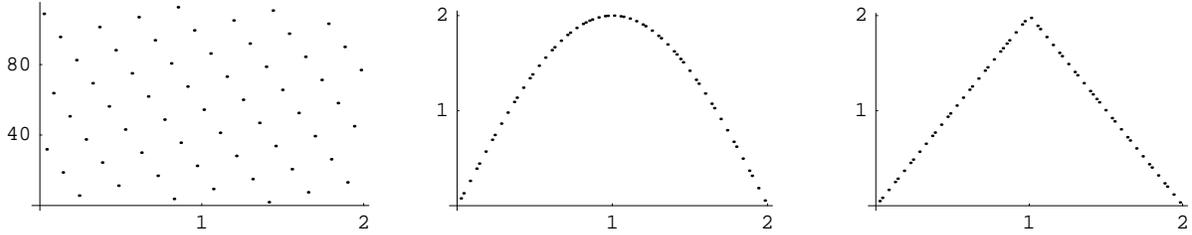}}}}
\caption{ \label{ds} $d_H(0,\phi_k)$, $d(0,\phi_k)$,
$d_E(0,\phi_k)$. Vertical unit is $\frac{\pi R}{\sin\theta\pi}$,
horizontal unit is $\pi$.}
\end{figure}

 In this sense, if we imagine an observer localized
at $\xox$ and whose only
information about the geometry of the surrounding world is the
measurement of the function $d(0,\phi)$, $S_x$ looks "smoother than a circle".
More rigourously, (\ref{dphi}) turns
out to be the length $L(\phi)$ of the minimal arc joining the
origin to a point $\phi$ on the cardioid with polar equation
\begin{equation}
\label{cardiobis} r = \frac{\pi\mathcal{R}}{4}(1+\cos\varphi).
\end{equation}
\begin{figure}[ht]
\begin{center}
\mbox{\rotatebox{0}{\scalebox{.55}{\includegraphics{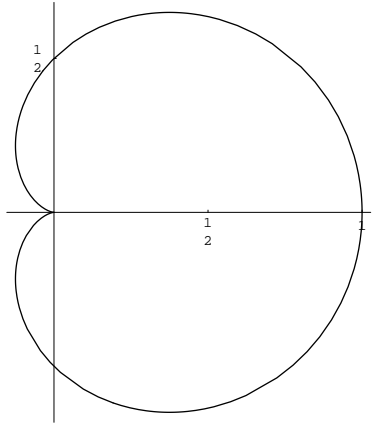}}}}
\end{center}
\caption{ \label{cardio} Cardioid $r =
\frac{\pi\mathcal{R}}{4}(1+\cos\varphi)$. Units are in $\frac{\pi
R}{\abs{\sin \theta\pi}}$.}
\end{figure}

Indeed restricting to $0\leq \phi\leq\pi$ (since $L(\phi) = L(2\pi
-\phi)$), \begin{equation} \label{dcardio} L(\phi) = \int_0^{\phi}
\sqrt{r^2 + (\frac{dr}{d\varphi})^2} d\varphi =
 \int_0^{\phi} \frac{\pi\mathcal{R}}{2}\cos \frac{\varphi}{2} d\varphi = \pi\mathcal{R}\sin\frac{\phi}{2} = d(0,\phi).
 \end{equation}

One has to be careful with the interpretation of equation
(\ref{dcardio}). The noncommutative geometry distance does {\it not}
turn the loop $S_x$ into a cardioid. What the noncommutative metric
does is to turn $S_x$ into an object that looks like a cardioid for
an observer localized at $x$ who is measuring the distance between
him and a point of $S_x$. Corollary \ref{dense} being invariant
under a re-parameterization of the basis $S^1$ ($\tau \rightarrow
\tau + \text{const.}$), the same analysis is true for an observer
localized at $y\neq x$. In this sense the cardioid point of view is
an intrinsic point of view.
\newline

\begin{figure}[ht]
\begin{center}
\mbox{\rotatebox{0}{\scalebox{.9}{\includegraphics{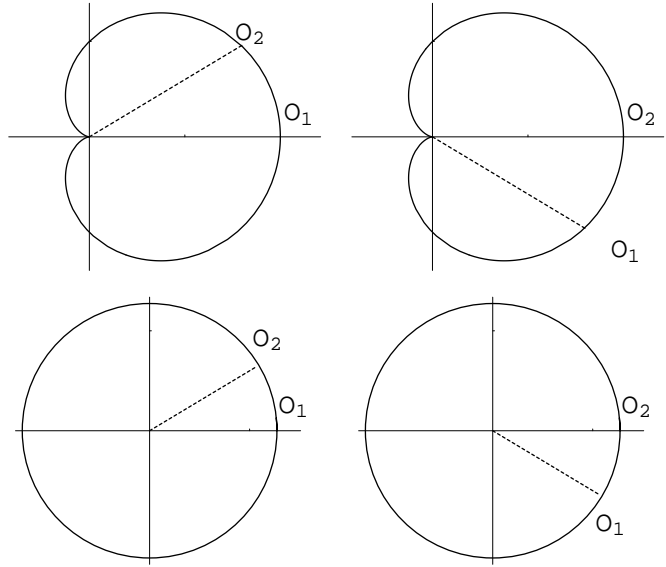}}}}
\end{center}\caption{ \label{according} On the left, the loop $S$ according to $\oo_1$; on
the right, the loop $S$ according to $\oo_2$. At bottom $S$ is  a
circle and one goes from left to right by re-parameterization. On
top $S$ is the fiber $S_x$ and a single manifold cannot encompass
both points of view.}
\end{figure}
 Things are
clearer in analogy with the circle (Figure \ref{according}):
consider $2$ observers $\oo_i$, $i=1,2$, located at distinct points
$\phi_i$ on a loop $S$. Assume each of them measures its own
distance function
$$d_i: z\in S \mapsto  d_i(x_i,z).$$
If both find that $d_i = d_E$, then they will agree that $S$ is  a circle.
On the contrary if both find that $d_i =
d$, then each of them will
pretend to be localized at the point
opposite to the cut locus of the cardioid and they will disagree on the nature
of $S$.
 In fact their disagreement is only due to their belief that $S$ is a manifold.
What the present work shows is precisely that the loop $S_x$
equipped with the noncommutative metric $d$ is {\it not} a manifold.
This example nicely illustrates how the distance formula
(\ref{distance}) allows one to define on very simple objects (like
tori) a metric which is not accessible from classical differential
geometry.
\newpage

\subsection{The shape of the basis}
\begin{figure}[hh]
\begin{center}
\mbox{\rotatebox{0}{\scalebox{.6}{\includegraphics{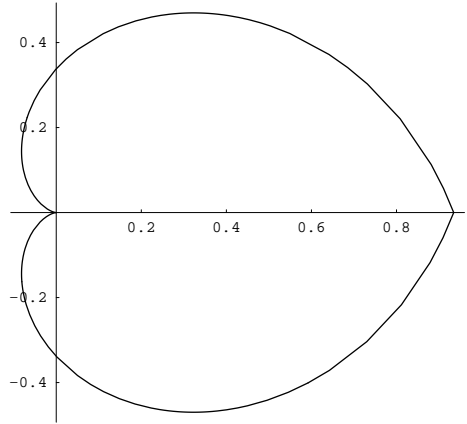}}}}
\end{center}\caption{ \label{coeur} The shape of $S_\xi$ when $\theta\rightarrow 1$.}
\end{figure}

From an intrinsic point of view the fiber looks like a cardioid.
What does the base $M=S^1$ look like ? Let $S_\xi$ denote the set of
points of $\mathbb{T}_\xi$ corresponding to the same vector $\xi\in
\cc P^{n-1}$,
$$S_\xi\doteq \{ p\in P,\; V(p)=\xi\}.
$$
We parameterize $S_\xi$ by
$\varphi\in[0,2\pi[$ with $\xox = 0$. Any point in $S_\xi$ can be obtained as a
$\zeta_\tau$ where $\tau = 2k\pi + \varphi$ and $\zeta$ defined by (\ref{zetay})
with
\begin{equation}
\label{thetabase}
\theta' = -\theta\tau.
\end{equation}

 In order to compute $d_H$, note that $\zeta_\tau$ is accessible from  $\xox$ if and only if
$\zeta_0$ is accessible, that is to say iff $\theta' = 2k'\theta\pi
\text{ mod } [2\pi]$ for some integer $k'$. In other words $
\text{Acc}(\xox)\cap S_\xi$ is the subset of $[0,2\pi[$ given by the
numbers $\varphi$ that write
$$ \varphi =  2p\theta^{-1}\pi + 2p'\pi$$
for some integers $p,p'$. When $\theta$ is irrational
$\text{Acc}(\xox)\cap S_\xi$ is dense in $S_\xi$ and to a given
$\varphi$ corresponds one and only one couple of integers $p,p'$. By
$(\ref{thetabase})$ one obtains
$$\zeta_0 = \xi_0^{-(k+p')},$$
where we used the notation (\ref{xitauk}). Hence $\zeta_\tau =
\xi_{2p\theta^{-1}\pi}$, so that
$$d_H(0,\varphi) \doteq d_H(\xox, \zeta_\tau) = 2p\theta^{-1}\pi.
$$
As in the case of the fiber $S_x$, one finds close to $0\in S_\xi$
in the Euclidean topology some points that are infinitely
Carnot-Carath\'eodory far from $0$. Hence $d_H$ not only forgets the
shape of the fiber but also the shape of the base.

On the contrary the noncommutative distance $d$ is finite on $S_\xi$
and preserves the shape of the base, although the latter is deformed
in a slightly more complicated way than the fiber. Note that, via
(\ref{thetabase}),
$$W_k= \frac{\abs{\sin (\frac{\theta}{2}\varphi)}}{\abs{\sin \theta\pi }}\,,\;\,
W_{k+1}= \frac{\abs{\sin (\frac{\theta}{2}(2\pi -\varphi))}}{\abs{\sin \theta\pi }}
$$
are independent of $k$. The same is true for $X$ and $Y$ so that
$d(0,\phi)= d(\xox,\zeta_\tau)$ only depends on $\varphi$ as
expected. Explicitly, defining $\lambda\doteq \frac{\varphi}{2\pi}$,
Proposition \ref{s1} writes
\begin{equation}
\label{carabasemoins}
\displaystyle
d(0,\varphi)= \pi\mathcal{R}\lp \lambda \sin (\theta\pi(1-\lambda)) +
(1- \lambda)\sin (\theta\pi\lambda)\rp
\end{equation}
when $Y$ is negative and
\begin{equation}
\label{carabaseplus}
d(0,\varphi)=   \left\{
\begin{array}{ll}
\displaystyle
2\pi\lp\mathcal{R}(\frac{1}{2}-\lambda)\sin \theta\pi\lambda + \lambda\rp
&\text{when }\lambda\leq \frac{1}{2}\\\displaystyle
2\pi\lp\mathcal{R}(\lambda -\frac{1}{2})\sin \theta\pi(1-\lambda) + 1- \lambda\rp
&\text{when }\lambda\geq \frac{1}{2}.\end{array}
\right.
\end{equation}
when $Y$ is positive. Even for a fixed value of $R$, $Y$ may change
sign when $\varphi$ runs from $0$ to $2\pi$ so it seems difficult to
find for $S_\xi$ a picture like the cardioid for $S_x$. However,
assuming that $Y$ is always negative, one can view the first line of
(\ref{carabasemoins}) as a kind of convex deformation of a cardioid.
In particular when $\theta\rightarrow 1$ or $\theta\rightarrow 0$,
$Y$ is indeed negative for any $\varphi$ so that
$$\underset{\theta\rightarrow 1}{\lim}\, d(0,\varphi) = \pi\mathcal{R}\sin \frac{\varphi}{2}$$
which corresponds to the length on a cardioid of infinite radius
(since $\underset{\theta\rightarrow 1}{\lim}\mathcal{R} = +\infty$),
while
$$\underset{\theta\rightarrow 0}{\lim}\, d(0,\phi) = 2R\varphi(1-\frac{\varphi}{2\pi}).$$
This is the arc length of the curve $r(\varphi)$, solution of
\begin{equation}
\label{equadiff}
r^2 + \dot{r}^2 = (1-\frac{\varphi}{\pi})^2.
\end{equation}
(\ref{equadiff}) has no global solution. Gluing the solution of
$\dot{r} = \sqrt{(1-\frac{\varphi}{\pi})^2 - r^2}$ on $[\pi,2\pi]$
with the solution of $\dot{r} = -\sqrt{(1-\frac{\varphi}{\pi})^2 -
r^2}$ on $[0,\pi]$ with initial condition $r(\pi) = 0$, one obtains
that at the limit $\theta\rightarrow 1$ the base $S_\xi$, seen for
$\xi$,  has the shape of a heart (figure \ref{coeur}). Hence, still
from the intrinsic point of view developed from $S_x$, $\theta$ is a
deformation parameter for the base of $P$ from an infinite cardioid
to a heart. The shape of $S_\xi$ for intermediate values of $\theta$
is deserving of further study.

\section{Conclusion and outlook}

The $2$-torus $\mathbb{T}_\xi$ inherits from noncommutative geometry
a metric smoother than the Euclidean one (the associated distance
function is smooth at the cut locus). It gives to both the fiber and
the base the shape of a cardioid or a heart. Such a "smooth
cardio-torus" (shall we denote it $\heartsuit_\xi$ ?) offers a
concrete example in which the distance (\ref{distance}) is "truly"
noncommutative, in the sense that is not a Riemannian geodesic
distance (as in the commutative case), nor a combination of the
latter with a discrete space (as in the two-sheet model), not even
the Carnot-Carath\'eodory one. The noncommutative distance combines
some aspects of the Euclidean metric on the torus (preservation of
the fiber structure) with some aspects of the Sub-Riemannian metric
(dependance on the connection).

From a geometrical point of view several questions remain to be
studied: what is the metric when both the scalar and the gauge
fluctuations are non-zero ? How to extend the present result to
manifolds other than $S^1$ ? In particular it could be interesting
to separate in the holonomy conditions the role of the curvature
from the role of the non-connectedness. For instance could it be
that, in a certain "local" sense, $d$ equals $d_H$ ? Let us also
underline that the present work is intended to be the first step in
the computation of the metric aspect of the noncommutative torus
where the bundle of pure states $P$ is no longer trivial.

From a physics point of view, it would be interesting to reexamine
in the light of the present results some interpretations that were
given to sub-Riemannian-geodesics as effective trajectories
 of particles (Wong's equations). This should be the object of further work.
 \newline

\noindent {\bf Acknowledgments} A preliminary version of Proposition
\ref{lemme1} was established by T. Krajewski. B. Iochum suggested to
study the example $M=S^1$, and pointed out that the conditions of
Corollary \ref{span} were not equivalent to the holonomy being
trivial. Thanks to all the NCG group of CPT and IML for numbers of
useful discussions, and to the director for hosting. Warm thanks to
P. Almeida and R. Montgomery for illuminating remarks. Many thanks
to A. Greenspoon for careful reading of the manuscript.
\newline

\noindent {\small{Work partially supported by EU network {\it
geometric analysis}. CPT is the UMR 6207 from CNRS and Universities
de Provence, M\'editerran\'ee and Sud Toulon-Var, affiliated to the
FRUMAM.}}

\end{document}